\documentclass[12pt]{article}
\usepackage{graphicx}
\usepackage{amssymb}
\begin{document} 

\begin{titlepage}
	\rightline{}
	
	
	\vskip 2cm 
	\begin{center}
		\Large{{\bf Scrambling with\\Matrix Black Holes}}
	\end{center}
	
	\vskip 2cm 
	\begin{center}
		{Lucas Brady\footnote{\texttt{lucas\_brady@hmc.edu}}\ \ \ and\ \ \ Vatche Sahakian\footnote{\texttt{sahakian@hmc.edu}}}\\
	\end{center}
	\vskip 12pt 
	\centerline{\sl Harvey Mudd College} 
	\centerline{\sl Physics Department, 241 Platt Blvd.}
	\centerline{\sl Claremont CA 91711 USA}
	
	\vskip 1cm 
	\begin{abstract}
		If black holes are not to be dreaded syncs of information but be fully described by unitary evolution, they must scramble in-falling data and eventually leak it through Hawking radiation. Sekino and Susskind have conjectured that black holes are fast scramblers: they generate entanglement at a remarkably efficient rate, with characteristic time scaling logarithmically with the entropy. In this work, we focus on Matrix theory -- M theory in the light-cone frame -- and directly probe the conjecture. We develop a concrete test bed for quantum gravity using the fermionic variables of Matrix theory and show that the problem becomes that of chains of qubits with an intricate network of interactions. We demonstrate that the black hole system evolves much like a Brownian quantum circuit, with strong indications that it is indeed a fast scrambler. We also analyze the Berenstein-Maldacena-Nastase model and reach the same tentative conclusion.
	\end{abstract}
\end{titlepage}

\newpage \setcounter{page}{1} 
\section{Introduction and Highlights}
\label{sub:intro}

That General Relativity breaks down near the singularity of a black hole is in principle well-established. And string theory promises to remedy this pathology with the help of an infinite tower of degrees of freedom and associated non-local dynamics (see for example~\cite{polchinski}). That this non-locality may even extend all the way to the black hole horizon is however still a tantalizing open question. There are no fundamental reasons to expect that spacetime near a black hole horizon -- where curvatures can be very small -- is described by anything but Einstein gravity. But then there remains the lingering issue of the information paradox~\cite{Hawking:1976ra,mathurinformation}. Black hole complementarity~\cite{Susskind:1993if,Lowe:1995ac,Kiem:1995iy} along with string theory address (see for example~\cite{Maldacena:2001kr,Balasubramanian:2011ur,Balasubramanian:2011dm}) -- at some level -- physics near the horizon: an in-falling observer falls past the horizon unscathed, but her information gets heavily entangled with the Hilbert space of the black hole -- we say the in-falling information is {\em scrambled} by the black hole. The external observer can then recover the in-falling data from entanglements with the Hawking radiation and there is no loss of information to worry about. The wrinkles in this narrative are still to be ironed out. More recently, there have been suggestions that this picture may be inconsistent. The most precise formulation of this challenge is through the firewall proposal of~\cite{Almheiri:2012rt}. In this latter scenario, the in-falling observer enjoys an even shorter lifespan and gets incinerated near the horizon -- even for a large black hole.

While the issues surrounding the firewall proposal are still being debated, in this work we go back to the original premise of black hole complementarity to directly tackle certain details. It was argued in~\cite{Sekino:2008he,Susskind:2011ap} that, if $\tau$ is the black hole {\em scrambling time} measured in the Schwarzschild time coordinate -- we must require a lower bound on $\tau$ 
\begin{equation}\label{eq:scramblingtime}
	\tau \geq \frac{\ln S}{T}
\end{equation}
where $T$ is the temperature of the black hole and $S$ is the entropy. Otherwise, one can devise a thought experiment that violates the no-cloning hypothesis of quantum mechanics. Knowing that a black hole is a thermal state, one expects that in-falling information is thermalized once it crosses the horizon. A thermalization process can involve several timescales. To address the no-cloning bound, the timescale of most relevance is the scrambling time $\tau$ which we need to define precisely: dividing up the whole black hole Hilbert space in two roughly equal parts, the scrambling time is the duration it takes for the two halves to equilibrate, to get fully entangled after starting from zero entanglement. More generally, scrambling time is the equilibration time of any subpart of the whole -- a subpart that occupies up to half of the whole. Most systems scramble information as a power law in entropy. Hence, black hole quantum cloning may be avoided by a safe margin. In~\cite{Sekino:2008he,Susskind:2011ap}, it was suggested and argued that black holes are however {\em fast scramblers} -- they are characterized by scrambling times of order $\ln S$ -- thus saturating the no-cloning safety bound~(\ref{eq:scramblingtime}). There is a certain elegance in this picture in that, the extreme objects that the black holes are, they test the limits of the laws of physics -- yet without violating them. This conjecture can be motivated by analyzing the diffusion rate of charge on a black hole horizon, or through Matrix theory~\cite{Banks:1996vh}, a framework of M theory in the light-cone gauge.

Matrix theory is an ideal setup for testing scrambling. It provides a concrete computational setting for quantum gravity -- one involving the dynamics of D0 branes. A Matrix black hole model was developed in~\cite{Horowitz:1997fr,Banks:1997hz,Banks:1997tn} showing that, under certain assumptions, the correct scaling of the equation of state of a Schwarzschild black hole can be reproduced within the theory. This proposal suggests modeling the black hole as a thermal configuration of D0 branes, strongly interacting through a dense network of strings stretching between them. However, the model needs a mechanism to treat the D0 branes as {\em distinguishable} particles to reproduce the correct entropy, and this in turn seems to require the presence of a background D2 brane structure -- perhaps a membrane stretched at the horizon. In short, while very promising, the Matrix black hole model remains incomplete with several missing key dynamical ingredients. More recently, \cite{Sekino:2008he} argued that the network of interactions between the bosonic degrees of freedom in Matrix theory may have the right structure and details to naturally lead to fast scrambling. Other attempts of understanding scrambling dynamics in Matrix and Matrix-like theories  can be found in~\cite{Asplund:2011qj}-\cite{Asplund:2012tg}.

A scrambling time scaling logarithmically with the entropy is a rather extreme situation. In~\cite{Lashkari:2011yi}, attempts were made to construct toy models of fast scramblers using techniques from the field of quantum information. Two particular models were presented with some promising attributes: a Brownian quantum circuit, and an Ising model on a random graph. Neither however was related to Matrix theory or to a theory of quantum gravity. 

In this work, we set up a concrete Matrix black hole model that provides a playground for testing scrambling in quantum gravity. We start with the bosonic variables of Matrix theory -- describing the coordinates of the D0 branes -- arranged in a fixed spherical shape, presumably the size of a would-be horizon. We do this by considering either Matrix theory in a box or the Berenstein-Maldacena-Nastase (BMN) model~\cite{Berenstein:2002jq} -- that is, Matrix theory in the background of a plane wave. In either case, this configuration is stabilized through a combination of background gravitational curvature and/or a background gauge flux providing a dielectric Myers effect~\cite{Myers:1999ps}. We then focus on fluctuations of this configuration, both bosonic and fermionic ones, to account for the thermodynamics. The fermionic excitations represent the $\mathcal{N}=1$ eleven dimensional supergravity multiplet -- {\em i.e.} low energy M-theory. And we focus on this fermionic sector to assess scrambling.

We show that the fermionic degrees of freedom organize themselves into spherical harmonic modes that can be represented by a network of qubit chains. If we ignore the effect of fluctuating bosonic degrees of freedom, we have qubit chains with a sparse nearest-neighbor network of interactions. For matrix theory, we deal with one dimensional chains; for the BMN model, two dimensional ones. For $U(N)$ Matrix or BMN theory, our problem consists of a Hilbert space of $8\,N^2$ total qubits, distributed amongst $N$ disconnected qubit chains. For Matrix theory, the longest chain has $2N-1$ qubits, for the BMN model it has $4N-2$ qubits. The interactions between different qubits within a chain effectively connect different modes of supergravity spherical harmonics -- transferring entanglement across the energy scales. And interactions between coarse modes are stronger than the ones between finer modes: long wavelengths dynamically pack more energy than shorter ones, reminiscent of stringy UV-IR mixing~\cite{Peet:1998wn}. However, this is a zeroth order computation: these qubit chains are in fact coupled to the fluctuations of the bosonic degrees of freedom. For quantum fluctuations of the bosonic background, we show that the qubit system becomes very much like the Brownian quantum circuit of~\cite{Lashkari:2011yi,dankert} -- a strong indication that Matrix theory is a fast scrambler. We also assess the effect of thermal fluctuations of the bosonic degrees of freedom. This leads to a hierarchy between two timescales: a fast time scaling as $1/T$ characterizing the bosonic dynamics, and a much slower one describing the fermionic dynamics. This is a classic setup for thermalization through a back-reaction mechanism. We show that the net effect is to create a dense network of interactions between the qubits -- interactions that go well-beyond the zeroth order nearest-neighbor pattern -- leading once again to an ideal setting for fast scrambling. 

In summary, we develop a concrete and controlled framework where we can use methods of quantum information theory in a theory of quantum gravity, and map Matrix theory dynamics onto more familiar qubit-qubit dynamics along models entertained by~\cite{Lashkari:2011yi}. All preliminary indications from this analysis point towards the conclusion that both Matrix theory and the BMN model are fast scramblers. To take the analysis to the next level of certainty, ones needs either numerical simulations or a limiting regime (such as large $N$) to simplify the otherwise complex dynamics. Hence, we also develop the framework to simulate these systems numerically. The setup turns out to be very challenging even for numerical methods! One needs to deal with immense Hilbert spaces even for $N\sim 10$. To do this, we develop original highly parallelized algorithms to simulate and assess scrambling. We show that we are able to evolve an initial state numerically and observe the development of entanglement across the Hilbert space of the system with enough statistics to yield robust conclusions. We use these methods to confirm that the zeroth order description yields power law scaling for scrambling time in terms of entropy, $\tau\sim S$ for Matrix theory and $\tau\sim S^{1/2}$ for the BMN model -- as expected for one and two dimensional nearest-neighbor qubit chains respectively. We defer work on simulating the full interacting system and the large $N$ analytical treatment to a future work~\cite{wip}.

The remaining discussion is organized as follows. In Section~\ref{sec:matrixfermions}, we derive the qubit chain Hamiltonians for Matrix theory and the BMN model; we analyze the Hilbert space and formulate the quantum information problem. In Section~\ref{sec:results}, we summarize the numerical techniques which include original methods in parallelizing the computation in manners that allow us to efficiently explore Hilbert spaces of up to $2^{23}$ dimensions. We present results from our simulations, including details of analysis, assessment of ergodicity and fluctuations at equilibrium, fitting, error estimation, and numerical error tracking. Finally, in Section~\ref{sec:conclusion}, we collect our thoughts and present future directions in pursuing the analysis. We also comment on the role of other thermalization timescales that we see in our system at earlier time steps.

\section{Matrix fermions on a sphere}\label{sec:matrixfermions}

\subsection{The Hamiltonian}

We start with the BMN Lagrangian~\cite{Berenstein:2002jq}
\begin{eqnarray}
	L&=&\frac{1}{2}g_s\mbox{Tr} \left[
	\frac{1}{g_s^2}\left(D_t X_i\right)^2+\frac{1}{g_s^2}\left(D_t Y_a\right)^2 + \frac{1}{2} \left[
	X_i, X_j\right]^2+\left[X_i, Y_a\right]^2+\frac{1}{2}\left[Y_a,Y_b\right]^2 \right. \nonumber \\
	&-& \left. \left(\frac{\mu}{3 g_s}\right)^2 X_i^2-\left(\frac{\mu}{6 g_s}\right)^2 Y_a^2-\frac{i}{3} \frac{\mu}{g_s}\, \epsilon_{ijk} \left[X_i,X_j\right] X_k \right. \nonumber \\
	&+& \left. \frac{1}{g_s}\Psi D_t \Psi + \Psi \gamma_i \left[X_i, \Psi\right]+\Psi \gamma_a \left[Y_a,\Psi\right]-\frac{i}{4} \frac{\mu}{g_s} \Psi \gamma_{123} \Psi
	\right]
\end{eqnarray}
in units where the eleven dimensional Planck length is one, $l^{(11)}_P=1$\footnote{To switch to string units, scale time as $t\rightarrow g_s^{-2/3} t$, and the matrices as $X\rightarrow g_s^{-1/3} X$ and $\Psi\rightarrow g_s^{-1/2} \Psi$.}. The $X_i$'s with $i=1,2,3$, and the $Y_a$'s with $a=1,\ldots, 6$, are $N\times N$ hermitian matrices; $\Psi$ is also an $N\times N$ hermitian matrix but its entries are Majorana-Weyl spinors in ten dimensions; $g_s$ is the string coupling; and $\mu$ is the BMN deformation parameter. In the limit $\mu\rightarrow 0$, we recover the Matrix theory Lagrangian of D0 branes in flat space. The covariant time derivative for the $U(N)$ gauge group involves the non-dynamical hermitian matrix $A$
\begin{equation}
	D_t = \partial_t-i\left[A,\,.\,\right]\ .
\end{equation}
The system has $16$ nontrivial supersymmetries; and an additional $16$ supersymmetries acting in the $U(1)$ sector of $U(N)$ -- describing the transformations of the center of mass degrees of freedom of the D0 branes.

We next fix a spherical configuration of D0 branes in the $X_1$-$X_2$-$X_3$ subspace, and analyze the dynamics of fluctuations on this sphere. We write the ansatz
\begin{equation}
	X_i=\nu \tau_i+x_i\ \ \ ,\ \ \ Y_a = y_a
\end{equation}
where the $\tau_i$'s satisfy the $SU(2)$ algebra
\begin{equation}
	\left[\tau_i,\tau_j\right] = 2 i \varepsilon_{ijk} \tau_k\ .
\end{equation}
and $\nu$ is some constant. The matrices $x_i$ and $y_a$ are taken as small. Without these perturbations, the configuration is interpreted as a fuzzy sphere of radius
\begin{equation}\label{eq:radius}
	R=\sqrt{\frac{\mbox{Tr} X_i^2}{N}} = \nu {\frac{\sqrt{N^2-1}}{\sqrt{3}}}\sim \nu N\ \ \ \mbox{for large $N\gg 1$}\ .
\end{equation}
It is a BPS state in the BMN theory if we choose
\begin{equation}
	\nu=\frac{\mu}{6\, g_s}\ .
\end{equation}
In the Matrix theory limit $\mu\rightarrow 0$, the spherical configuration is unstable - albeit with a long lifetime for large $N$~\cite{Sahakian:2001zs}. One can stabilize it by adding a mass term to the Lagrangian of the form $m^2\mbox{Tr} X^2$; physically, this is achieved by putting the D0 branes in a non-trivial supergravity background where $m^2$ is mapped onto local tidal forces experienced by the D0 branes~\cite{Sahakian:2000bg}\footnote{In particular, in a term of the form $m_{ij} X_i X_j$, the mass parameter maps onto $m_{ij}\propto g_{tt,ij}$ where $g_{tt}$ is the time-time component of a background metric.}.

For simplicity, we imagine the background compactified to $3+1$ dimensions: we do this by freezing the dynamics in the $Y_a$ directions, setting $y_a=0$ by hand. Including this six dimensional dynamics does not present any conceptual complications. 

The next step is to decompose the matrix structure of $\Psi$, $A$, and $x_i$ in a basis of spherical harmonics matrices $Y^{j}_{m}$~\cite{Dasgupta:2002hx,hoppe,reza}
\begin{equation}
	\Psi_\alpha=\psi^j_{m\,\alpha} Y^j_m\ \ \ ,\ \ \ 
	A = a^j_m Y^j_m\ \ \ ,\ \ \ 
	x_i = x^j_{m\,i} Y^j_m\ ,
\end{equation}
where we have explicitly added the spinor index $\alpha$ on $\psi^j_m$, with $\alpha=1,\ldots, 16$; while the $a^j_m$'s and $x^j_{m\,i}$'s are bosonic variables. The $Y^{j}_{m}$'s are $N\times N$ matrices and span the space of hermitian matrices for $j=0,\ldots, N-1$, and $m=-j\cdots j$. They satisfy an orthogonality condition
\begin{equation}
	\mbox{Tr} \left(
	Y^j_m Y^{j'}_{m'}
	\right) = (-1)^m N \delta_{jj'} \delta_{-mm'}\ ;
\end{equation}
and the conjugation identity
\begin{equation}
	(Y^{j}_{m})^\dagger = (-1)^m Y^{j}_{-m}\ .
\end{equation}
This implies that -- given that $\Psi$ is a hermitian matrix of Majorana-Weyl spinors -- we must have
\begin{equation}\label{eq:daggerpsi}
	(\psi^j_{m\,\alpha})^\dagger = (-1)^m \psi^j_{-m\,\alpha}\ .
\end{equation}
Similarly, we have
\begin{equation}\label{eq:daggerx}
	(x^j_{m\,i})^\dagger = (-1)^m x^j_{-m\,i}\ .
\end{equation}
The commutation relations of the $Y^{j}_{m}$'s with $SU(2)$ generators are given by
\begin{equation}
	\left[\tau_+,Y^{j}_{m}\right] = \sqrt{(j-m)(j+m+1)} Y^{j}_{m+1}
\end{equation}
\begin{equation}
	\left[\tau_-,Y^{j}_{m}\right] = \sqrt{(j+m)(j-m+1)} Y^{j}_{m-1}
\end{equation}
\begin{equation}
	\left[\tau_3,Y^{j}_{m}\right] = m Y^{j}_{m}
\end{equation}
where we define
\begin{equation}
	\tau_1 = \tau_++\tau_-\ \ \ ,\ \ \ 
	\tau_2 = -i(\tau_+-\tau_-)
\end{equation}
More generally, we have the algebra~\cite{hoppe,reza}
\begin{equation}
	\left[
		Y^{j}_{m},Y^{j'}_{m'}
	\right] = f_{jm,j'm',j''m''} (-1)^{m''} Y^{j''}_{-m''}
\end{equation}
where $f_{jm,j'm',j''m''}$ is constructed from $3j$ amd $6j$ symbols as follows
\begin{eqnarray}
	&& f_{jm,j'm',j''m''} = \frac{2}{N}(-1)^N N^{3/2} \sqrt{(2j+1)(2j'+1)(2j''+1)} \times \\
	&&
	\left(
	\begin{array}{ccc}
		j & j' & j'' \\
		m & m' & m''
	\end{array}
	\right)\times
	\left\{
	\begin{array}{ccc}
		j & j' & j'' \\
		\frac{N-1}{2} & \frac{N-1}{2} & \frac{N-1}{2}
	\end{array}
	\right\}
\end{eqnarray}
if $j+j'+j''$ is odd; otherwise $f_{jm,j'm',j''m''}=0$.
Note in particular that
\begin{equation}
	\tau_+= \sqrt{\frac{2}{N}} Y^{1}_{1}\ \ \ ,\ \ \ 
	\tau_-= \sqrt{\frac{2}{N}} Y^{1}_{-1}\ \ \ ,\ \ \ 
	\tau_3= \sqrt{\frac{2}{N}} Y^{1}_{0}\ \ \ . 
\end{equation}

Using these expressions, we can now write the Lagrangian in terms of the $\psi^{j}_{m\,\alpha}$, $x^{j}_{m\,i}$, and $a^j_m$. The latter is a Lagrange multiplier whose equations of motion immediately yield the $N^2$ Gauss law constraints
\begin{equation}\label{eq:constraint}
	\sum_{j,m}\sum_{j',m'}f_{jm,j'm',j''m''} \psi^{j}_{m\,\alpha} \psi^{j'}_{m'\,\alpha} = 0\ ,
\end{equation}
with the conventional gauge choice $a^j_{m}=0$ (corresponding to $A=0$). Tracing over the matrix structure entirely, the Hamiltonian splits into several pieces which we write as
\begin{equation}
	H=H_x + H_{x,\psi} + H_\psi
\end{equation}
where $H_x$ is the part independent of the fermionic variables, $H_{x,\psi}$ contains the terms coupling the $x$'s and $\psi$'s, and the $H_\psi$ is the rest.
We start by focusing on the last piece $H_\psi$. The gauge fixed form becomes
\begin{eqnarray}
	H_\psi &=& N\, g_s\,\left[ \nu\frac{1}{4} \sum_{j=0}^{N-1}\sum_{m=-j}^j (-1)^m \sqrt{1+j-m} \sqrt{j+m} \psi^{j}_{1-m} \gamma^1 \psi^{j}_{m} \right. \nonumber \\
	&+& \nu\frac{1}{4} \sum_{j=0}^{N-1}\sum_{m=-j}^j (-1)^m \sqrt{1+j+m} \sqrt{j-m} \psi^{j}_{-1-m} \gamma^1 \psi^{j}_{m} \nonumber \\
	&+& \nu\frac{i}{4} \sum_{j=0}^{N-1}\sum_{m=-j}^j (-1)^m \sqrt{1+j-m} \sqrt{j+m} \psi^{j}_{1-m} \gamma^2 \psi^{j}_{m} \nonumber \\
	&-& \nu\frac{i}{4} \sum_{j=0}^{N-1}\sum_{m=-j}^j (-1)^m \sqrt{1+j+m} \sqrt{j-m} \psi^{j}_{-1-m} \gamma^2 \psi^{j}_{m} \nonumber \\ 
	&-& \nu\frac{1}{2} \sum_{j=0}^{N-1}\sum_{m=-j}^j (-1)^m m \psi^{j}_{-m} \gamma^3 \psi^{j}_{m} \nonumber \\
	&+& \left.\frac{i}{8} \left(\frac{\mu}{g_s}\right) \sum_{j=0}^{N-1}\sum_{m=-j}^j (-1)^m \psi^{j}_{-m} \gamma^{123} \psi^{j}_{m}\right]\ . \label{eq:mainham}
\end{eqnarray}
The momentum canonical to $\psi^{j}_{m}$ is
\begin{equation}
	\Pi^{j}_{m\, \alpha} = \frac{1}{2} (-1)^m N \psi^{j}_{-m\,\alpha}
\end{equation}
yielding the anticommutation relations
\begin{equation}\label{eq:anticomm}
	\left\{
	\psi^{j}_{m\, \alpha}, (\psi^{j'}_{m'\, \alpha'})^\dagger
	\right\}  = \frac{2}{N} \delta_{\alpha\alpha'} \delta_{jj'} \delta_{mm'}\ .
\end{equation}
We will first analyze this system, ignoring the bosonic fluctuations, then come back to the full problem.

\subsection{Hilbert space}

Introducing rescaled variables for $m>0$
\begin{equation}
	s^{j}_{m\, \alpha} \equiv \sqrt{\frac{N}{2}} \psi^{j}_{m\, \alpha}\ \ \ \mbox{for $m>0$}\ ,
\end{equation} 
equation~(\ref{eq:anticomm}) becomes
\begin{eqnarray}
	&& \left\{
	s^{j}_{m\, \alpha}, (s^{j'}_{m'\, \alpha'})^\dagger
	\right\}  = \delta_{\alpha\alpha'} \delta_{jj'} \delta_{mm'}\ \ ,\ \ \left\{
	s^{j}_{m\, \alpha}, s^{j'}_{m'\, \alpha'}
	\right\}  = 0 \ \ ,\ \ \left\{
	(s^{j}_{m\, \alpha})^\dagger, (s^{j'}_{m'\, \alpha'})^\dagger
	\right\}  = 0 \nonumber \\
	&& \mbox{for $m=1\cdots j$,\ \ \ \   $j=1\cdots N-1$,\ \ \ \   and $\alpha=1\cdots 16$}\ ;
\end{eqnarray}
this is the canonical creation/annihilation algebra for a collection of fermions.
The $m=0$ case is special since $(\psi^{j}_{0\, \alpha})^\dagger = \psi^{j}_{0\, \alpha}$ from~(\ref{eq:daggerpsi}), which leads to
\begin{equation}
	\left\{
	\psi^{j}_{0\, \alpha}, \psi^{j'}_{0\, \alpha'}
	\right\}  = \frac{2}{N} \delta_{\alpha\alpha'} \delta_{jj'}\ .
\end{equation}
We then define
\begin{equation}
	\Gamma_{j\, \alpha}\equiv \sqrt{N} \psi^{j}_{0\, \alpha} \Rightarrow
	\left\{
		\Gamma_{j\,\alpha}, \Gamma_{j'\,\alpha'}\right\} = 2\, \delta_{\alpha\alpha'} \delta_{jj'}\ .
\end{equation}
Hence, we instead get a Clifford algebra in the $m=0$ sector.

We can now analyze the Hilbert space. For fixed $j$ and $m>0$, we have $16\times j$ qubits that can be created and destroyed with $(\psi^{j}_{m\, \alpha})^\dagger$ and $\psi^{j}_{m\, \alpha}$. With $j$ ranging from $0$ to $N-1$, this gives $8\,N\,(N-1)$ qubits and hence a $256^{N\,(N-1)}$ dimensional Hilbert space. There are also the $m=0$ modes that yield a Clifford representation in $16\times N$ dimensions (that is, a representation of $spin(16N)$). This means we have an additional $8N$ qubits arising from the $m=0$ modes and this a $256^N$ dimensional representation. In total, we then have a system of $8\,N\,(N-1)+8N=8N^2$ qubits, and a Hilbert space of $256^{N^2}$ dimensions. This maps onto the $256$ states of
the $\mathcal{N}=1$ $D=10+1$ supergravity multiplet (that is, low energy M-theory) -- raised to the dimensionality of the $U(N)$ group. Hence, our system corresponds to the dynamics of $256$ gravitons, gravitinos, and 3-form gauge particles on a sphere: the $N^2$ modes are the harmonics or Kaluza-Klein modes on the fuzzy sphere which comes with a UV cutoff due to its discretized matrix structure.

We can build up the Hilbert space of the system by applying raising operators on a state $|\Omega\rangle$ annihilated by all lowering operators. To define $|\Omega\rangle$, we then write
\begin{equation}\label{eq:vac1}
	s^{j}_{m\, \alpha} |\Omega\rangle = 0\ \ \ \mbox{for $m=1\cdots j$,\ \  $j=1\cdots N-1$,\ and $\alpha=1\cdots 16$}\ .
\end{equation}
and
\begin{equation}\label{eq:vac2}
	\Gamma^-_{j\, \alpha} |\Omega\rangle = 0\ \ \ \mbox{for $j=1\cdots N-1$,\ and $\alpha=1\cdots 8$}\ ,
\end{equation}
where we have defined
\begin{equation}
	\Gamma^+_{j\,\alpha} \equiv \frac{1}{2} \left(
	\Gamma_{j\,\alpha} + i \Gamma_{j\, \alpha+8}
	\right)\ \ \ ,\ \ \ \Gamma^-_{j\,\alpha} \equiv \frac{1}{2} \left(
	\Gamma_{j\, \alpha} - i \Gamma_{j\, \alpha+8}
	\right) 
\end{equation}
with $\alpha=1,\ldots,8$ and corresponding anticommutation relations
\begin{equation}
	\left\{\Gamma^+_{j\, \alpha}, \Gamma^-_{j'\, \alpha'}\right\} = \delta_{\alpha\alpha'} \delta_{jj'}\ \ \ ,\ \ \ 
		\left\{\Gamma^+_{j\, \alpha}, \Gamma^+_{j'\, \alpha'}\right\} = 0\ \ \ ,\ \ \ 
		\left\{\Gamma^-_{j\, \alpha}, \Gamma^-_{j'\, \alpha'}\right\} = 0
\end{equation}
In doing so, we have chosen a particular structure of the Clifford representation that combines the $\alpha$'th spinor with the $\alpha+8$'th spinor. In tune with this, we will shortly choose a representation of the gamma matrices appearing in the Hamiltonian that will make the interaction pattern amongst the qubits more transparent. With the definition of $|\Omega\rangle$ given by~(\ref{eq:vac1}) and~(\ref{eq:vac2}), any state in the Hilbert space is then generated by acting upon $|\Omega\rangle$ by a number of $(s^j_{m\,\alpha})^\dagger$'s with $m>0$ and $1\leq\alpha\leq 16$, and $\Gamma^+_{j\,\alpha}$'s with $1\leq\alpha\leq 8$. Each action of a raising operator flips a corresponding qubit on; and applied more than once, it kills the state. We thus generate the entire $256^{N^2}$ states of the Hilbert space as all possible configurations of $8\,N^2$ qubits.

\subsection{Interactions}

\begin{figure}
	\begin{center}
		\includegraphics[width=6.5in]{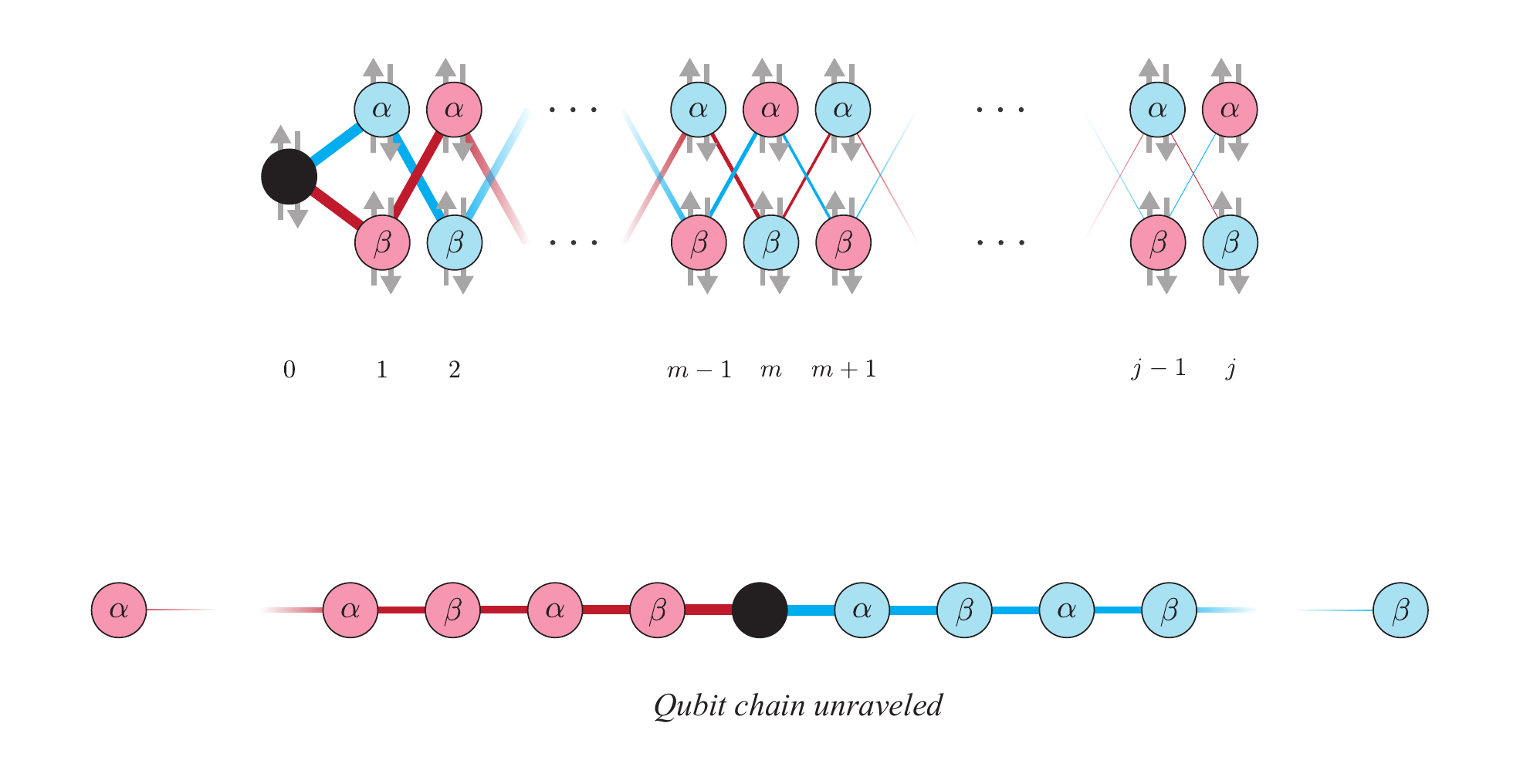} 
	\end{center}
	\caption{A graphical representation of the Matrix theory qubit chain. Each circle represents a qubit site -- labeled by a spinor index $\alpha=1,\ldots,8$ or $\beta=\alpha+8$, and a spherical harmonic mode $m=1,\ldots,j$. In addition, the $m=0$ site is shown as a black qubit on the left for a total of $2\,j+1$ qubits in each fixed-$j$ sector. The lines connecting the qubits represent interactions: thicker lines imply stronger interactions, thinner ones weaker. We see that we have two chains; and the two chains are connected to each other through the $m=0$ qubit.}\label{fig:mat}
\end{figure}

We now go back to equation~(\ref{eq:mainham}) to analyze the structure of qubit-qubit interactions. At this stage, it helps to fix a representation of the gamma matrices appearing in the Hamiltonian: $\gamma_1$, $\gamma_2$, and $\gamma_3$. We collect the details of this in the Appendix. Once a representation is chosen, we expand and simplify the resulting Hamiltonian and we write the end results separately for the Matrix theory ($\mu\rightarrow 0$) case, and the BMN case ($\mu\neq 0$). The Matrix model yields a simpler Hamiltonian
\begin{eqnarray}
	H_\psi &=& N g_s\sum_{j=0}^{N-1} \sum_{m=0}^j \sum_{\alpha=1}^8 
	\left[ 2\,m\, (\psi^j_{m\,\alpha})^\dagger \psi^j_{m\,\alpha} - 2\,m\, (\psi^j_{m\,\beta})^\dagger \psi_{\beta,j,m} \right. \nonumber \\
	&+& \left. i \sqrt{(j-m)(j+m+1)} \left(
	\psi^j_{m\,\alpha} (\psi^j_{m+1\,\beta})^\dagger + (\psi^j_{m\,\beta}) (\psi^j_{m+1\,\alpha})^\dagger \right.\right. \nonumber\\
	&+& \left.\left. (\psi^j_{m\,\alpha})^\dagger \psi^j_{m+1\,\beta} + (\psi^j_{m\,\beta})^\dagger (\psi^j_{m+1\,\alpha}) 
	\right) \right] \label{eq:matrixH}
\end{eqnarray}
where $\beta\equiv \alpha+8$. Note in particular that the sum over $\alpha$ ranges now from $1$ to $8$ only, and the qubit-qubit interactions involve couplings between each $\alpha$th qubit with the one at $\beta=\alpha+8$. For $m=0$, this is why the natural representation for the Clifford algebra -- within this chosen representation of the $\gamma$ matrices -- pairs the $\alpha$th and the $\alpha+8$th matrices of $spin(16N)$. Staring at~(\ref{eq:matrixH}), we see that we have chains of qubits interacting with nearest neighbor interactions. We also notice that sectors of qubits with different $j$ do not interact. Hence, we can focus on a fixed-$j$ sector in analyzing the time evolution of an initial state. As mentioned above, the interactions also do not mix different spinor indices $\alpha=1,\ldots,8$: for a fixed $\alpha$, a chain mixes only a fixed pair of spinor indices $\alpha$ and $ \beta=\alpha+8$. These facts greatly simplify the analysis. However, we also note that the coefficients of the interactions are not independent of $m$, which makes the model significantly harder to tackle than, for example, a standard Hubbard model. To see the problem at hand more transparently, we can reparameterize the spinors as follows
\begin{equation}
	\Sigma^{(1)}_m\equiv \sqrt{\frac{N}{2}}\left\{
	\begin{array}{ll}
		\psi_{m\,\alpha} & \mbox{for odd $m$} \\ 
		\psi_{m\,\beta} & \mbox{for even $m$}
	\end{array}\right. \ \ \ ,\ \ \ 
	\Sigma^{(2)}_m\equiv \sqrt{\frac{N}{2}}\left\{
	\begin{array}{ll}
		\psi_{m\,\beta} & \mbox{for odd $m$} \\ 
		\psi_{m\,\alpha} & \mbox{for even $m$}
	\end{array}\right.
\end{equation} 
where we have dropped the $j$ label since we can focus on a fixed $j$ sector at a time. The Hamiltonian in a fixed $j$ and $(\alpha,\beta)$ sector then takes the form
\begin{eqnarray}
	H^j_\alpha&=&-i g_s \sum_{m=1}^j\sqrt{(j-m)(j+m+1)} \left[ (\Sigma^{(1)}_{m+1})^\dagger \Sigma^{(1)}_m+\Sigma^{(1)}_{m+1}(\Sigma^{(1)}_m)^\dagger+(\Sigma^{(2)}_{m+1})^\dagger\Sigma^{(2)}_m+\Sigma^{(2)}_{m+1}(\Sigma^{(2)}_m)^\dagger
	\right] \nonumber \\
	&-& \sum_{m=1}^j m \left[ (\Sigma^{(1)}_m)^\dagger \Sigma^{(1)}_m-\Sigma^{(1)}_m(\Sigma^{(1)}_m)^\dagger \right]
	+ \sum_{m=1}^j m \left[ (\Sigma^{(2)}_m)^\dagger \Sigma^{(2)}_m-\Sigma^{(2)}_m (\Sigma^{(2)}_m)^\dagger \right] \nonumber \\
	&-& i \sqrt{j (j+1)} \left[ 
	-(\Sigma^{(2)}_1)^\dagger\Sigma^{(2)}_0-\Sigma^{(2)}_0\Sigma^{(2)}_1+(\Sigma^{(1)}_1)^\dagger\Sigma^{(1)}_0-\Sigma^{(1)}_0\Sigma^{(1)}_1\right] \nonumber \\
	&-& i \sqrt{2 j} \left[ (\Sigma^{(2)}_{j-1})^\dagger\Sigma^{(2)}_j - (\Sigma^{(2)}_j)^\dagger\Sigma^{(2)}_{j-1}
	 \right]
\end{eqnarray}
The full Hamiltonian is
\begin{equation}
	H_\psi = \sum_{j=0}^{N-1} \sum_{\alpha=1}^8 H^j_\alpha\ .
\end{equation}
For fixed $j$ and $\alpha$, we then have two chains with $j$ qubits each -- one consisting of the $\Sigma^{(1)}$ spinors, the other of the ${\Sigma^{(2)}}$'s -- and each with nearest neighbor interactions with $m$ dependent interaction coefficients. In addition, one has one additional qubit site corresponding to the $m=0$ mode represented by
\begin{equation}
	\Sigma^{(1)}_0 = \frac{i}{\sqrt{2}} \left( \Gamma^--\Gamma^+ \right)\ \ \ ,\ \ \ 
	\Sigma^{(2)}_0 = \frac{1}{\sqrt{2}} \left( \Gamma^-+\Gamma^+ \right)\ .
\end{equation}
Thus, the two chains interact with the $m=0$ qubit. In summary, we effectively have a single chain with $2j+1$ qubits, with site dependent interactions. Figure~\ref{fig:mat} depicts a cartoon of this setup, along with the corresponding $\alpha$, $\beta$, and $m$ labels. 

The site dependent interactions in our system make an integrable structure, if any, far from obvious. We have tried unsuccessfully to diagonalize this Hamiltonian through analytical methods. And diagonalizing it numerically, we have preliminary evidence that -- while the eigenvalue spectrum can be determined exactly -- the structure of eigenvectors hints at chaotic dynamics. We will report on this in a separate work~\cite{wip}
\footnote{An alternative approach is to first diagonalize the interactions in the Hamiltonian, which is straightforward. Introducing a redefinition of our spinor variables as in 
\begin{equation}
	S^j_{m\,\alpha}\equiv (-1)^\alpha \sqrt{\frac{j+m+1}{2j+1}} \psi^j_{m\,\alpha}+\sqrt{\frac{j-m}{2j+1}}\psi^j_{m+1\,\beta}\ \ \ ,\ \ \ 
	T^{j}_{m\,\alpha}\equiv -(-1)^\alpha \sqrt{\frac{j-m}{2j+1}} \psi^j_{m\,\alpha}+\sqrt{\frac{j+m+1}{2j+1}}\psi^j_{m+1\,\beta}\ ,
\end{equation}
the Hamiltonian then becomes
\begin{equation}
	H_\psi = -\frac{N}{2} \sum_{j=0}^{N-1} \sum_{m=1}^j \sum_{\alpha=1}^8 j\, (S^j_{m\,\alpha})^\dagger S^j_{m\,\alpha} + \frac{N}{2} \sum_{j=0}^{N-1} \sum_{m=1}^j \sum_{\alpha=1}^8 (j+1)\, (T^{j}_{m\,\alpha})^\dagger T^{j}_{m\,\alpha}
\end{equation}
with new raising/lowering operators $S^\dagger$/$T^\dagger$ and $S$/$T$. However, the new $S$ and $T$ variables have  relations amongst them because of the condition $(\psi^j_{m\,\alpha})^\dagger = (-1)^m \psi_{\alpha\,j-m}$. Alternatively, only a subspace of the Hilbert space built with these new variables is physical, and determining this subspace involves solving a set of complicated constraint equations. At the end, we find that this approach is more computationally involved than the one adopted in the text.  
}.

\begin{figure}
	\begin{center}
		\includegraphics[width=6.5in]{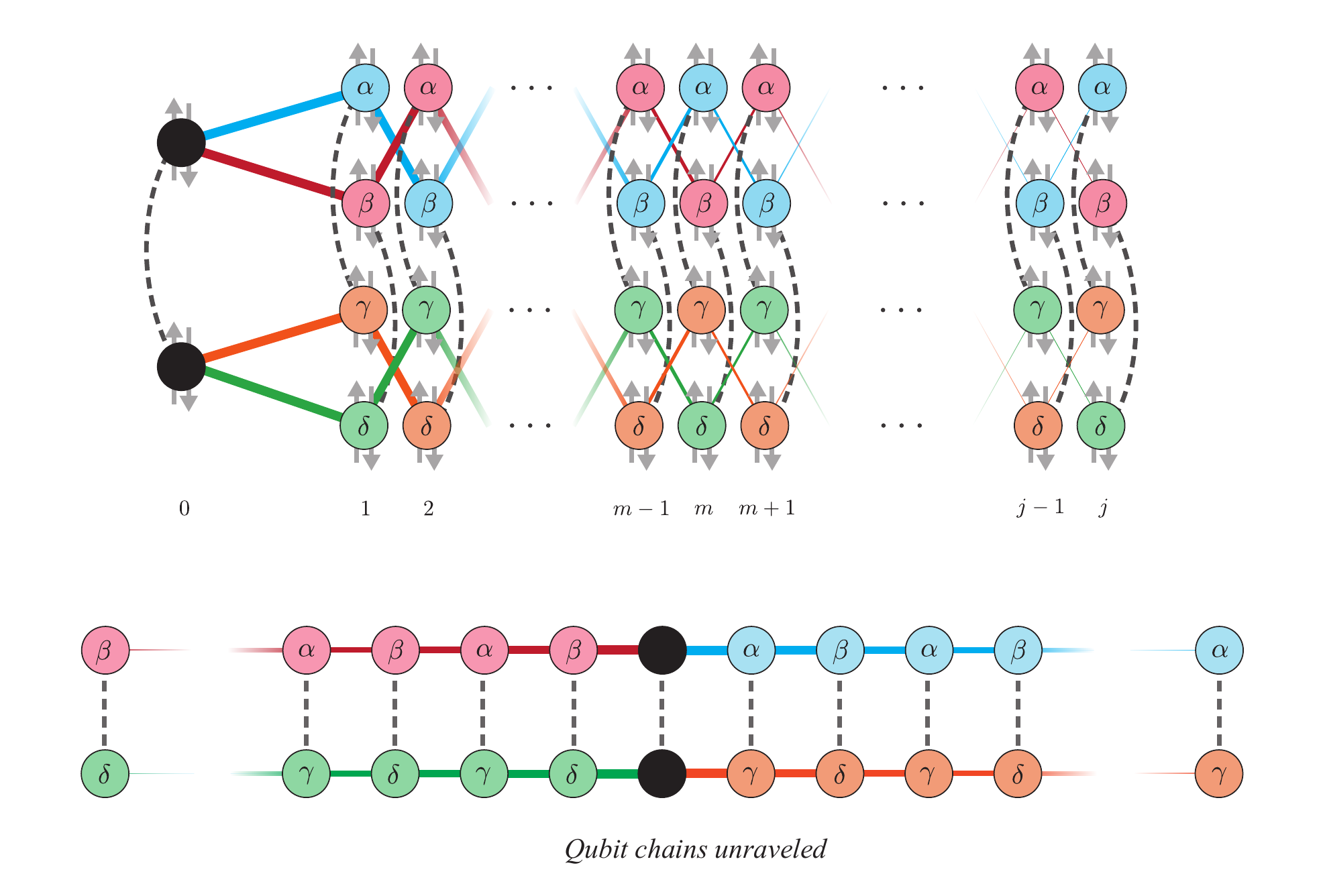} 
	\end{center}
	\caption{A graphical representation of the BMN chains. Each circle represents a qubit site -- labeled by a spinor index $\alpha=1,\ldots,4$, $\beta=\alpha+8$, $\gamma=\alpha+4$, and $\delta=\alpha+12$, and a spherical harmonic mode $m=1,\ldots,j$. In addition, the $m=0$ sites are shown as black qubits on the left for a total of $4\,j+2$ qubits in each fixed-$j$ sector. The lines connecting the qubits represent interactions: thicker lines imply stronger interactions, thinner ones weaker. We see that we have four chains; and these four chains are connected to each other through the two $m=0$ qubits. In addition, the dashed lines represent the new BMN interactions weighted by $\mu$. The result is then a two dimensional array of interacting qubits.}\label{fig:bmn}
\end{figure}

Repeating the same analysis for the BMN case, one obtains, after some lengthy but straightforward algebra, the Hamiltonian
\begin{eqnarray}
	H_\psi &=& N g_s\sum_{j=0}^{N-1} \sum_{m=0}^j \sum_{\alpha=1}^4 \left[  2m (\psi^j_{m\,\alpha})^\dagger \psi^j_{m\,\alpha} - 2\,m (\psi^j_{m\,\beta})^\dagger \psi^j_{m\,\beta} \right.  \\*
	&+& 2\,m (\psi^j_{m\,\gamma})^\dagger \psi^j_{m\,\gamma} - 2\,m (\psi^j_{m\,\delta})^\dagger \psi^j_{m\,\delta} \\*
	&+& i \left(\frac{\mu}{g_s}\right) (-1)^\alpha \left(
	\psi^j_{m\,\gamma} (\psi^j_{m\,\alpha})^\dagger - \psi^j_{m\,\alpha} (\psi^j_{m\,\gamma})^\dagger
	\right) \nonumber \\
	&+& i \left(\frac{\mu}{g_s}\right) (-1)^\beta \left(
	\psi^j_{m\,\delta} (\psi^j_{m\,\beta})^\dagger - \psi^j_{m\,\beta} (\psi^j_{m\,\delta})^\dagger
	\right) \\*
	&+& \left. i \sqrt{j+m+1}\sqrt{j-m} \left(
	\psi^j_{m\,\alpha} (\psi^j_{m+1\,\beta})^\dagger + \psi^j_{m\,\beta} (\psi^j_{m+1\,\alpha})^\dagger \right.\right. \\*
	&+& (\psi^j_{m\,\alpha})^\dagger \psi^j_{m+1\,\beta} + (\psi^j_{m\,\beta})^\dagger \psi^j_{m+1\,\alpha}
	+ (\psi^j_{m\gamma})^\dagger \psi^j_{m+1\,\delta} + (\psi^j_{m\delta})^\dagger \psi^j_{m+1\,\gamma} \\*
	&+& \left.\left. (\psi^j_{m\,\gamma})^\dagger \psi^j_{m+1\,\delta} + (\psi^j_{m\,\delta})^\dagger \psi^j_{m+1\,\gamma}
	\right) \right]\postdisplaypenalty=100
\end{eqnarray}
where $\beta\equiv \alpha+8$, $\gamma\equiv \alpha+4$, and $\delta\equiv \alpha+12$; and now the range of $\alpha$ is $1,\ldots,4$. Once again this Hamiltonian splits into non-interacting sectors and can be written as
\begin{equation}
	H_\psi = \sum_{j=0}^{N-1}\sum_{\alpha=1}^4 H^j_\alpha\ .
\end{equation}
We summarize the result neatly in Figure~\ref{fig:bmn}. Within each $H^j_\alpha$, we can reparametrize the spinor degrees of freedom to make the structure of chains more transparent -- as we did in the Matrix theory case.  We then see that one has four chains interacting with each other through a network of connections and two $m=0$ modes -- for a total of $4\,j+2$ qubits. The additional BMN interactions between the qubits then lead to a full two dimensional array of links.

\subsection{Bosonic fluctuations}

The bosonic fluctuations yield the Hamiltonian
\begin{eqnarray}
	H_x&=& N\, g_s\,\left[ \nu^4 - \frac{1}{3} \left(\frac{\mu}{g_s}\right) \nu^3 + \frac{1}{18} \left(\frac{\mu}{g_s}\right)^2 \nu^2
	\right]\nonumber \\
	&+&\frac{1}{2} N g_s (-1)^m {\dot{x}}^{j}_{-m\,i} {\dot{x}}^{j}_{m\,i} \nonumber \\
	&+&\frac{\sqrt{N} g_s}{3} \nu \left( \mu-3\nu \right) \left[ 
	\sqrt{2} i f_{jm,j'm',10} x^{j}_{m\,2} x^{j'}_{m'\,1} \right. \nonumber \\
	&+&\left.\left(  f_{jm,j'm',1-1} - f_{jm,j'm',11}\right) x^{j}_{m\,3} x^{j'}_{m'\,1}
	+i \left( f_{jm,j'm',1-1} + f_{jm,j'm',11} \right) x^{j}_{m\ 3} x^{j'}_{m'\,2} 
	 \right] \nonumber \\
	&-& \frac{N g_s}{18} (-1)^m \left[ 
	\frac{1}{2} \left( 2\mu^2+9\nu^2(j+j^2+m^2) \right) \left( x^{j}_{-m\,1} x^{j}_{m\,1} + x^{j}_{-m\,2} x^{j}_{m\,2} \right) \right. \nonumber \\
	&+&\left. \left( \mu^2 +9 \nu^2  (j+j^2-m^2) \right) x^{j}_{-m\,3} x^{j}_{m\,3}
	+ \frac{i}{6} \left( 4\mu-3\nu \right) \nu m x^{j}_{m\,1} x^{j}_{-m\,2} 
	\right] \nonumber \\
	&+& \frac{N g_s}{2} \nu^2 \sqrt{(j-m)} \sqrt{j+m+1} (-1)^m \left[  
	- i\,x^{j}_{-m-2\,2} x^{j}_{m\,1} - i x^{j}_{-m-2\,1} x^{j}_{m\,2} \right. \nonumber \\
	&+&\left. m\,x^{j}_{-m-1\,3} x^{j}_{m\,1}
	- i\,m\, x^{j}_{-m-1\,3} x^{j}_{m\,2} \right. \nonumber \\
	&+& \left. \frac{1}{4} \sqrt{2+j+m} \sqrt{j-m-1} \left( x^{j}_{-m-2\,1} x^{j}_{m\,1} - x^{j}_{-m-2\,2} x^{j}_{m\,2} \right) + \mbox{c.c.}
	 \right]
\end{eqnarray}
where we have dropped cubic and quartic terms for simplicity. These latter terms are however important for getting the equation of state of the Matrix black hole scale correctly~\cite{Horowitz:1997fr,Banks:1997hz,Banks:1997tn}. We then have $3N^2$ oscillators, with a complex spectrum of frequencies, strongly interacting amongst themselves. 

\subsection{Coupling between qubits and bosonic fluctuations}\label{sec:coupling}

The last piece of our Hamiltonian takes a deceptively simple form
\begin{equation}\label{eq:coupling}
	H_{x,\psi}=\frac{1}{2} N g_s f_{jm,j'm',j''m''} x^{j}_{m\,i} \psi^{j''}_{m''}\gamma^i \psi^{j'}_{m'}\ .
\end{equation}
This term couples the bosonic fluctuations to the fermionic ones, the latter being organized into qubit chains earlier. The full system is a complex animal involving all these degrees of freedom interacting and evolving in a coupled manner. In Section~\ref{sec:results}, we present the results of numerical simulations in the absence of the coupling term~(\ref{eq:coupling}). These results confirm that, in the absence of the coupling term~(\ref{eq:coupling}), the qubit chain system is {\em not} a fast scrambler; instead, we see the expected power law scaling with entropy for one and two dimensional nearest-neighbor chains.

\begin{figure}
	\begin{center}
		\includegraphics[width=6.5in]{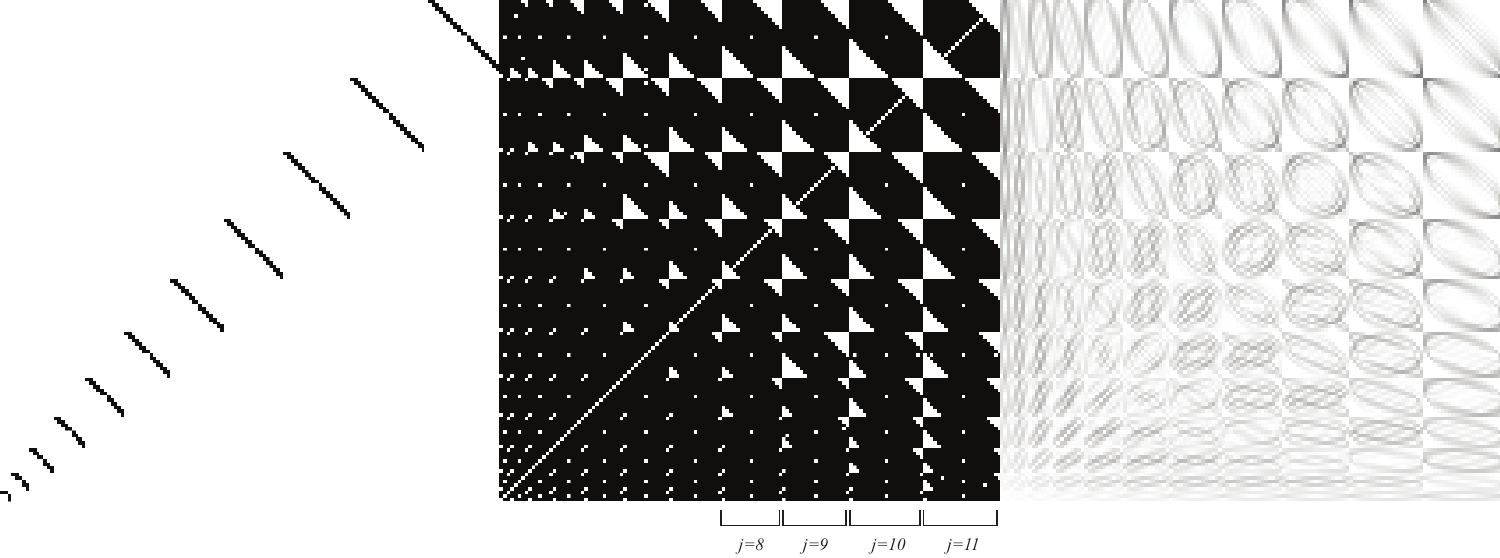} 
	\end{center}
	\caption{Graphs for links in qubit chains. The horizontal and vertical axes label qubit harmonics $(j,m)= (0,0)$,$(1,-1)$,$(1,0)$,$(1,1)$,$(2,-1)$, etcetera. On the left, we have the nearest-neighbor case; in the middle and the right, the Matrix or BMN case. In the graphs on the left and middle, black denotes a link between corresponding qubits, while white denotes no link. In the graph on the right, gray level indicates the strength of the links. We see that the number of links in the nearest-neighbor scenario scales as $N$. For the Matrix or BMN case, we see that a qubit chain with fixed $j$ connects with all other harmonics of different $j$'s. The network of new connections scales as $N^2$ as is obvious from the graph. Note that each block spans $m=-j,\ldots, j$ whereas independent qubits are labeled by $m=0,\ldots,j$ -- hence the folding symmetry of the pattern.}\label{fig:chains}
\end{figure}

We can gauge the effect of the coupling on the evolution of the qubit chains as follows. Let us imagine we start with the background bosonic fluctuations in a vacuum configuration and arrange an initial condition where the qubit chains are in the $|\Omega\rangle$ state\footnote{One can easily check that $|\Omega\rangle$ is within the physical space of states of the theory as discussed later.}. The fluctuations in the background would generically have a quantum-driven Gaussian profile. We can write
\begin{equation}
	\mbox{Prob} [x]= \sim e^{-\frac{x^2}{2 \sigma_x^2}}
\end{equation}
where the $\sigma_x^2$ can in principle be determined by studying correlations functions in the strongly coupled dynamics of the background fluctuations. The coupling~(\ref{eq:coupling}) then introduces new links between the qubits, making the interaction network denser while being weighted by a random Gaussian noise. We see that our system is then starting to look very much like the Brownian circuit discussed in~\cite{Lashkari:2011yi} which exhibits fast scrambling! There are however some differences. The Brownian circuit of~\cite{Lashkari:2011yi} connects every qubit to every other qubit with no correlations between the strengths of the links. This is not the case for our model. To estimate the density of the new links one generates through the bosonic quantum fluctuations, we write 
\begin{equation}
	H_{x,\psi}\sim  N g_s f_{jm,j'm',j''m''} x^{j}_{m} \psi^{j''}_{m''}(\gamma^1+\gamma^2+\gamma^3) \psi^{j'}_{m'}
\end{equation}
dropping the $i$ index since we expect isotropy on average. The gamma matrix structure adds links between the chains but this is of no relevant consequence: we start with $16$ possible qubit chains in spinor index space, the Hamiltonian $H_{\psi}$ connects these so that we have $8$ or $4$ independent chains for Matrix theory or the BMN model respectively, and finally this coupling term adds more connections that leaves us with $4$ independent chains in spinor index space (see Appendix A for the explicit form of $\gamma^1+\gamma^2+\gamma^3$). All this results in a number of additional links that does not depend on $N$; instead, it creates a number of links within the spinor space which necessarily will be of order one with respect to $N$. Hence, we can simplify things further by writing
\begin{equation}
	H_{x,\psi}\sim  N g_s f_{jm,j'm',j''m''} x^{j}_{m} \psi^{j''}_{m''}\psi^{j'}_{m'}
\end{equation}
dropping the spinor indices as well. The key then is the density of additional links generated by the coupling of various spherical harmonics modes through the coefficient $f_{jm,j'm',j''m''}$. Assuming random fluctuations in $x^{j}_{m}$ with no correlations with $j$ or $m$ for the sake of simplicity, we then have new links between the qubits at $j'm'$ and $j''m''$ if
\begin{equation}\label{eq:fjm}
	\sum_{j,m} f_{jm,j'm',j''m''} \neq 0\ .
\end{equation} 
If these coefficients are nonzero and uniform for {\em all} $j'm'$ and $j''m''$, we can then show that Matrix theory realizes a quantum Brownian circuit and hence leads to fast scrambling. The easiest way to address this question is to visualize these coefficients. Figure~\ref{fig:chains} shows three graphs corresponding to a nearest-neighbor network and the network generated by~(\ref{eq:fjm}).
We immediately see that the quantum fluctuations of the background bosonic coordinates do add a very large number of new links between the qubits, a number scaling as $N^2$, strongly suggesting the potential for fast scrambling. But this is not identical to the Brownian circuit: there are still missing links in the graph. In fact, the pattern is more interesting if we graph the strength of the links generated by~(\ref{eq:fjm}), as shown on the right in Figure~\ref{fig:chains}. A careful inspection indicates that shorter wavelengths couple stronger than longer ones, the opposite pattern seen in the zeroth order nearest-neighbor analysis. The devil is in the details, and to definitively conclude that the theory is a fast scrambler, we need further analysis, which we defer to~\cite{wip}.

We can also setup an alternative scenario to test fast scrambling. Imagine we start with a configuration of a Matrix black hole model -- a soup of strongly interacting bosonic oscillators and qubits -- all fluctuations of the fuzzy sphere the size of the would-be horizon. This soup is left to settle in a {\em thermal} state with temperature $T$, the temperature of the Matrix black hole. We would want the size of the black hole to be large, $R\sim N\gg 1$, implying that $T\sim 1/R \ll 1$. We expect that the strong coupling yields a robust thermal state; {\em i.e.} one that is not easily thrown out of equilibrium. The typical timescale of local thermalization in this system should be $1/T$~\cite{reifbook}. 

We then arrange the qubits of one particular chain of fixed $j\sim N$ into the pure state $|\Omega\rangle_j$ and let them evolve -- as if these are our probes of the thermal background. We want to find how long it takes for the initial state $|\Omega\rangle_j$ to scramble. This is an interesting example of quantum stochastic evolution of a qubit system with a fluctuating coupling~(\ref{eq:coupling}). Deferring a complete analytical and numerical treatment to a future work, we proceed with a heuristic analysis that clarifies many of the qualitative physical features.

The scrambling timescale of interest is necessarily much longer than $1/T$, if entropy $S\gg 1$, irrespective of whether our qubit chain is a fast scrambler or not. Hence, the dynamics we want to track has a timescale of evolution much longer than the timescale characterizing the background thermal state. Furthermore, the probe consists of about $N$ qubits, whereas the background has many more degrees of freedom -- scaling as $N^2$. We then propose that the thermal background functions like a heat reservoir. The strong coupling from the cubic and quartic terms in the Hamiltonian quickly dissipates, on a timescale of order $1/T$, any impact the qubit chain has on the thermal bath through~(\ref{eq:coupling}). But it is well-known that such an effect on a reservoir is not entirely negligible and is, in fact, crucial for the mechanism of thermalizing a probe~\cite{reifbook}: the qubit chain slightly alters the thermal reservoir, which in turn {\em back-reacts} on the dynamics of the chain through~(\ref{eq:coupling}), generating dissipation. As we have set things up, this process is necessarily Markovian\footnote{For fast scrambling of the probe, this feature is weakened but not invalidated as long as entropy $S\gg 1$. However, the same analysis can also be done in a more general setting involving non-Markovian thermalization at the cost of yielding integro-differential equations of evolution.}. A similar situation was numerically analyzed in Matrix theory in~\cite{Riggins:2012qt} where both probe and bath resided in the bosonic sector of the theory and the analysis was entirely classical. In that work, it was found that the scrambling timescale -- to the crude extent scrambling could be defined classically -- scaled as $1/T$, with no dependence on the entropy. \cite{Riggins:2012qt} proposed that this was a pathology of the classical treatment, and that a quantum mechanical analysis is probably needed to reveal the $\ln S/T$ dependence, if present. 

In the current context, we have a probe that is treated quantum mechanically -- a chain of qubits -- evolving in the fluctuating background of a thermal bath. Let $\delta t_0\sim 1/T$ denote the timescale characterizing the thermal background evolution. For example, for the bosonic fluctuations, we may write
\begin{equation}
	\langle x^j_{m\,i}(t)x^{j'}_{m'\,i'}(t+\delta t_0)\rangle_{th} \sim e^{-T \delta t_0}\ ,
\end{equation}
where the thermal averaging is equivalent to time averaging over timescales of order $\delta t_0$.
The timescale associated with the probe is denoted instead as $\delta t\gg \delta t_0$. To zeroth order in the back-reaction effect, we have
\begin{equation}
	\langle x^j_{m\,i}(t) \rangle_{th} \rightarrow 0 \Rightarrow \langle H_{x,\phi} \rangle_{th} = 0
\end{equation}
and we reach the incorrect conclusion that the system is not a fast scrambler.  However, the interaction with the probe shifts the bath energy slightly by an amount determined by equation~(\ref{eq:coupling}) so that the probability distribution becomes
\begin{equation}\label{eq:shiftx}
	\mbox{Prob} [x]= \sim e^{-\frac{x^2}{2 \sigma_x^2}} e^{-\beta \delta E_{x,\psi}}
\end{equation}
where $\sigma_x^2$ is now computed in a thermal bath, $\delta E_{x,\psi}$ is the change in the energy of the thermal bath over a short timescale $\delta t_0$ due to its interaction with the probe, and $\beta=1/T$. Introducing the operator
\begin{equation}
	\mathcal{O}^j_{m\,i}\equiv f_{jm,j'm',j''m''} \psi^{j''}_{m''}\gamma^i \psi^{j'}_{m'}\ ,
\end{equation}
we then have 
\begin{equation}
	\delta E_{x,\psi} \sim N g_s x^{j}_{m\,i} \langle \left[ H_\psi, \mathcal{O}^j_{m\,i}\right]\rangle \delta t_0
\end{equation}
where we take the expectation value of the spinor operators in the evolving qubit state. The bottom line is that this shifts the thermal average $\langle x^j_{m\,i}(t) \rangle_{th}$ through~(\ref{eq:shiftx}) to a non-zero value
\begin{equation}\label{eq:shift}
	\langle x^j_{m\,i}(t) \rangle_{th} \sim N g_s \beta \delta t_0 \sigma_x^2 \langle \left[ H_\psi, \mathcal{O}^j_{m\,i}\right]\rangle\ .
\end{equation}
Hence, the term~(\ref{eq:coupling}) gets turned on for timescales of order $\delta t$.
Once again, we add a dense network of connections between the qubits of our probe chain, along the pattern shown in Figure~\ref{fig:chains}, now weighted by thermodynamic parameters of the thermal bath through~(\ref{eq:shift}) -- leading to very promising prospects for fast scrambling.

\section{Simulations at zeroth order}\label{sec:results}

In this section, we develop new numerical techniques to analyze the problem at hand and apply them to the scenario where the coupling term~(\ref{eq:coupling}) is artificially turned off. We demonstrate that under these conditions, both Matrix theory and the BMN model scramble information as normal nearest-neighbor qubit chains do -- indicating that fast scrambling, if present, must arise through the coupling term~(\ref{eq:coupling}).

\subsection{Scrambling time}

We want to evolve an initial state of qubits with no coupling to the bosonic degrees of freedom, and track entanglement as a function of time.
In the Matrix theory case, we have a one dimensional chain with $q=2\,j+1$ qubits; in the BMN case, we have a two-dimensional network with $q=4\,j+2$ qubits. We choose the initial state as $| \Phi(t=0) \rangle_{j,\alpha} = |\Omega\rangle$ which is not an energy eigenstate. Given that the Hamiltonian does not mix between different $j$ and $\alpha$ sectors, we label the state with the appropriate subscripts. In choosing the initial state, we need to solve the constraint~(\ref{eq:constraint}) to assure that we are evolving within a physical subspace of the Hilbert space. The constraint is essentially the generator of $U(N)$ transformations (up to normal ordering). One can easily show that the expectation value of the constraint with respect to $|\Omega\rangle$ vanishes, implying that $|\Omega\rangle$ is a physical $U(N)$ invariant state. The constraint is then equivalent to the statement that a physical state must be constructed by applying on $|\Omega\rangle$ combinations of the spinor matrix $\Psi$ in a $U(N)$-invariant combination; for example, states of the form $\mbox{Tr} \Psi_\alpha \Psi_\beta \cdots |\Omega\rangle$ would be physical. We also know that the Hamiltonian evolution preserves the constraint once the initial state is chosen properly. For our purposes, we will then simply choose the initial state as the $U(N)$-invariant vacuum $|\Omega\rangle$ to keep things simple.
To probe the theories in an as ergodic regime as possible, we would be interested in the large $j$ case to explore a larger Hilbert space: in particular, we are interested in the $j=N-1$ sector. Hence, restricting to $j=N-1$, we start with $|\Omega\rangle$ in a chain of $q=2(N-1)+1=2 N-1$ qubits (Matrix case) or $q=4(N-1)+2=4 N-2 = 2(2 N-1)$ qubits (BMN case) -- ignoring all other $j$ sectors of the system; and then we evolve this state within the $j=N-1$ sector.

Once the initial pure state is chosen, its unitary evolution would be tracked by
\begin{equation}\label{eq:schrodinger}
	i \hbar \frac{\partial}{\partial t} | \Phi(t) \rangle_{N-1,\alpha} = H^{N-1}_{\alpha} | \Phi(t) \rangle_{N-1,\alpha}\ .
\end{equation}
We then identify a subset of qubits, henceforth denoted by $\mathcal{M}$, which is to become the ``system'' of interest. And we trace over the complement qubits $\overline{\mathcal{M}}$ to generate a density matrix $\rho(t)$ -- which experiences a non-unitary evolution and hence information loss. We write
\begin{equation}
	\rho(t) = \mbox{Tr}_{\overline{\mathcal{M}}}\,\, | \Phi(t) \rangle_{N-1,\alpha}\,\, \mbox{}_{N-1,\alpha}\langle \Phi(t) \rangle |\ .
\end{equation}
From this, we compute the von Neumann entropy
\begin{equation}
	S(t) = -\rho(t)\, \mbox{Tr}_\mathcal{M} \rho(t)\ .
\end{equation}
to assess entanglement and information scrambling and identify the relevant timescale. The goal is to determine this scrambling time as a function of the parameters of the problem, in particular as a function of the sizes of ${\mathcal{M}}$ and $\overline{\mathcal{M}}$. We accord to the definition that scrambling time is the time it takes for two halves of the whole to equilibrate\footnote{If this Hilbert space would be that of a black hole, this scrambling timescale would correspond to the minimum amount of time required for information about the initial state to be imprinted onto the Hawking radiation and thus leak out of the black hole.}. For a typical system with entropy $S$ at temperature $T$, the scrambling timescales as
\begin{equation}
	\tau\sim \frac{S^C}{T}
\end{equation}
for some constant $C$. A {\em fast scrambler} however scrambles in time $\tau\sim \ln S/T$.

\begin{figure}
	\begin{center}
		\includegraphics[width=6.5in]{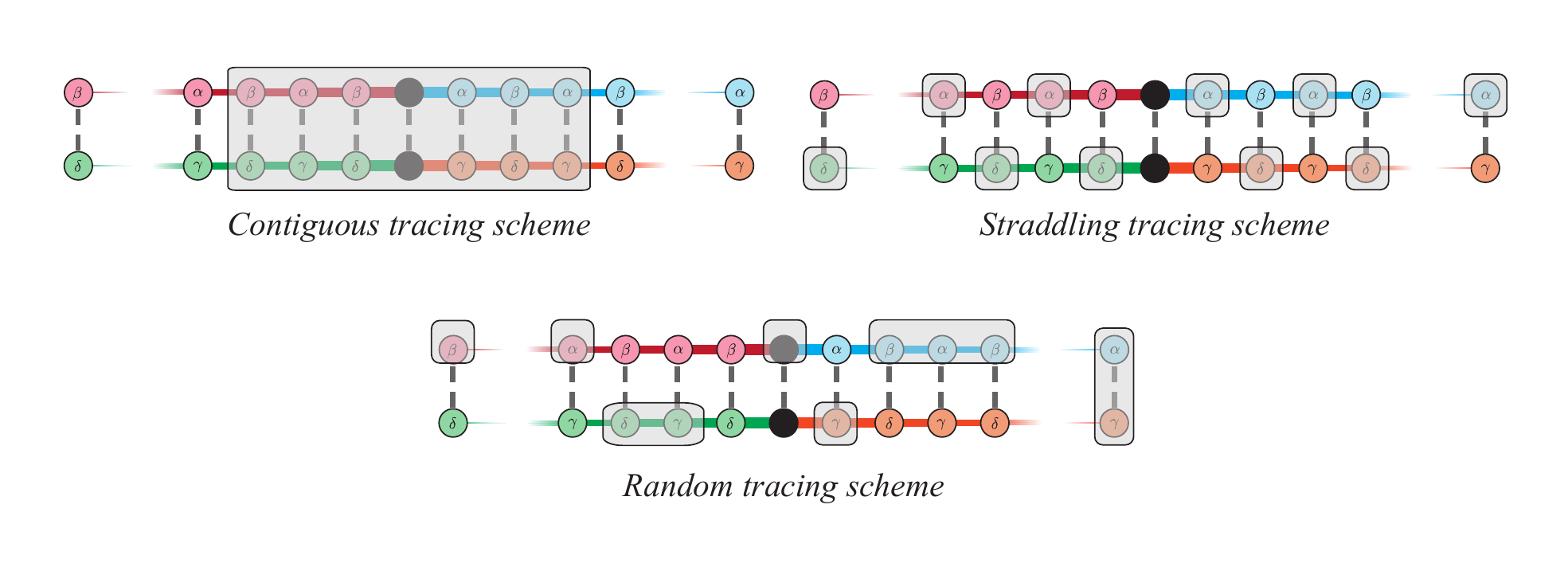} 
	\end{center}
	\caption{The three different tracing schemes we use to measure entanglement (BMN case depicted). The grayed out areas represent the qubits over which we trace ($\overline{\mathcal{M}}$) while the rest represent the system $\mathcal{M}$: (a) shows the case where we trace over a contiguous chunk of the qubit chains; this minimizes the interactions between $\mathcal{M}$ and $\overline{\mathcal{M}}$ expecting to lead to the least amount of entanglement growth; (b) depicts the case where we trace over a straddled subset that maximizes the connections between qubits in $\mathcal{M}$ and those in $\overline{\mathcal{M}}$, leading to more von Neumann entropy; finally (c) shows the case where we trace over a random set of qubits selected from the entire set $\overline{\mathcal{M}}+\mathcal{M}$; this is to represent a `generic' scenario.}\label{fig:tracing}
\end{figure}

It is known from quantum information theory that several factors affect entropy and scrambling time: the choice of initial state, the choice of the set $\mathcal{M}$ that constitutes the system, and the details of the qubit-qubit interactions. Our goal is to find out whether the latter attribute -- the details of qubit interactions -- leads to fast scrambling in Matrix or BMN theories. We choose $|\Omega\rangle$ as the initial state, making sure that it is not an eigenstate of the Hamiltonian. For the set of qubits $\mathcal{M}$, we consider several situations depicted in Figure~\ref{fig:tracing}: (1) a contiguous set along the qubit chains, which reduces the network of interactions between $\mathcal{M}$
and $\overline{\mathcal{M}}$ to the few qubits at their common boundaries; (2) a sequence of straddling qubits that maximizes the interaction links between $\mathcal{M}$ and $\overline{\mathcal{M}}$; (3) samples of qubits chosen randomly from the network chain to constitute the system $\mathcal{M}$. We also arrange that the number of qubits in $\mathcal{M}$ is around half of the total number of qubits in $\mathcal{M}+\overline{\mathcal{M}}$ -- while we vary this total number over a range. This is so as to explore a regime of scrambling most relevant to the black hole no-cloning argument of~\cite{Hayden:2007cs,Sekino:2008he,Susskind:2011ap}.

\subsection{Numerical methods}\label{sec:numerical}

For fixed $N$, $j=N-1$, and arbitrary fixed $\alpha$, we have a closed system with $2\,N-1$ qubits for Matrix theory or $4\,N-2$ qubits for the BMN case. This corresponds to a Hilbert space that is respectively $2^{2\,N-1}$ or $2^{4\,N-2}$ dimensional. Larger values of $N$ would lead to more succinct scrambling:  for smaller $N$ the setup would involve larger equilibrium fluctuations. In the case of Matrix theory, we find $N\sim 6$ has enough states in the Hilbert space so that the non-linear dynamics of the theory leads to scrambling. Note that, for $N\sim 10$, one is already dealing with a Hilbert space with a million states! 
This is a rather challenging problem even for numerical methods. We have two strategy options:
\begin{itemize}
	\item Numerically diagonalize the Hamiltonian in some basis; then use the energy eigenstates to evolve the initial state. This is technically a very computationally intensive numerical problem because it requires computation of eigenstates for a Hilbert space of very large dimension. The task becomes computationally prohibitive for $N>8$ for Matrix theory or $N>4$ for the BMN case.
	\item Numerically evolve the initial state over discretized time using a Runge-Kutta algorithm and equation~(\ref{eq:schrodinger}). Albeit still dealing with a large Hilbert space, this problem is comparatively tractable. 
\end{itemize} 

We then adopt the second method. It allows us to readily explore dynamics with $N\leq 12$ in Matrix theory or $N\leq 6$ in the BMN model. We employ the basis of states identified by the vacuum~(\ref{eq:vac1})-(\ref{eq:vac2}) and the corresponding raising and lowering operators $(\psi^j_{m\,\alpha})^\dagger$ and $\psi^j_{m\,\alpha}$. We represent the quantum state as a product of qubits, each qubit occupying only {\em a single bit} of computer memory to maximize efficiency. The Hamiltonian is then a sum of direct products of two by two matrices. Each step of the computation is then cast into the form of bit-shift operations -- a maximally efficient computational scheme. The algebra involves single precision arithmetics ($4$ bytes per float).
For the numerical evolution of the initial $|\Omega\rangle$ state, we employ a 4th order Runge Kutta algorithm to assure numerical stability. We use a small enough time step of $10^{-3}$ in dimensionless units ($l^{(11)}_P=1$) to control the accumulation of numerical errors. All this is still a very
involved computation that can only be reasonably tackled with a highly parallelized algorithm. We employ $896$ nVidia GPU cores (Tesla architecture), allowing us to explore Matrix theory with up to $N = 12$ -- or $23$ qubits -- in a time frame less than an hour per an entire simulation sequence. The main limitation of the setup becomes memory driven: we are not able to go beyond $N=12$ since this requires more than the $12 \mbox{GBytes}$ of GPU memory.

At every time step, we compute the density matrix by implementing a highly efficient original algorithm that reconstructs from the bits in memory representing the product state a reduced density matrix -- once again employing parallel processing. From the density matrix, we compute the von Neumann entropy, once again using GPU parallel processing. Entropy as a function of time is then the output from the simulation. This data is analyzed with a combination of Mathematica and Igor Pro. For every set of fixed parameters, several simulations are performed and analyzed to allow for error estimation. In addition, we assess numerical errors by computing the trace of the density matrix -- which is expected to remain equal to one as the time evolution progresses. Throughout our simulation, we find that the trace of the density matrix remains fixed at unity to {\em five significant digits} demonstrating that we effectively have no numerical errors to worry about. 

Without the highly threaded algorithm design we implemented - distributing the computation across hundreds of GPU cores - it would have been impossible to perform this computation with large enough $N$ to demonstrate scrambling. In this sense, the technical computational aspects of this project break or make the results.

\subsection{Analysis techniques and errors}

We run a sequence of simulations for the evolution of the qubit chains {\em without} the coupling term~(\ref{eq:coupling}). This is so as to test the numerical techniques and confirm that our system is not a fast scrambler when not taking into account coupling to bosonic degrees of freedom.
 
The main output from our simulations is the von Neumann entropy as a function of simulation time step. In total, we present results from the following simulations:

\begin{itemize}
	\item Matrix theory with $N=3$ to $N=12$ ($5, 7, 9, \ldots, 23$ total qubits, and system sizes ranging from $2, 3, 4, \ldots, 11$ qubits respectively)
	
	\begin{itemize}
		\item with the straddled tracing scheme.
		\item with a random tracing scheme.
		\item with the contiguous tracing scheme.
	\end{itemize}
		
	\item BMN model with $N=2, 3, 4, 5, 6$ ($6, 10, 14, 18, 22$ total qubits, and system sizes $2, 4, 6, 8, 10$ qubits respectively)
	
	\begin{itemize}
		\item with the straddled tracing scheme.
		\item with a random tracing scheme.
		\item with the contiguous tracing scheme.
	\end{itemize}
	
\end{itemize}

	In addition, we present one tangential case study: we arrange a setup in Matrix theory with a system of $3$ qubits and a whole of $20$ qubits to attempt to capture a reservoir-like setup with the whole significantly bigger than the system. We use this simulation to elaborate on other timescales involved in the equilibration process.
	
	For all simulations, we fix the initial state to $|\Omega\rangle$ as described earlier -- a state that lives in the physical Hilbert space of the corresponding theory.
	
	To quantify the rate of scrambling, we adopt the following criterion. We divide up the qubits of the system into two sets with an approximate $40$-$60$ ratio. The slightly smaller subspace is then viewed as our ``system''. The system and the rest interact and thus exchange energy as they equilibrate; the total energy, which is conserved, is fixed by the choice of an initial pure state. After evolving the full pure state, we trace over the qubits that are not in our system, generating a density matrix. The von Newmann entropy computed from this density matrix is then a measure of the level of entanglement between our system and the rest. To identify scrambling time, we look for the timescale at which the von Neumann entropy $S$ of our system levels off near its equilibrium value $S_{eq}$, which is necessarily smaller than the maximum possible entropy $S_{max}=q\ln 2$ for a system of $q$ qubits. Note that our system is {\em not} much smaller than the whole as is more common in reservoir-based setups\footnote{In a reservoir setup, the von Neumann entropy also corresponds to the the measure of thermodynamic entropy. We are extending this notion to our setup of two equally sized systems exchanging energy with each other since an assessment of entanglement should correspond to an assessment of the onset of ergodicity in phase space.}. From this arrangement, we then extract the scrambling timescale $\tau$ and the equilibrium entropy $S_{eq}$. And we do all this as we vary system size -- the number of qubits or equivalently $N$ while maintaining the $40$-$60$ proportion -- looking at the dependence of the scrambling timescale and equilibrium entropy on the size of the Hilbert space of our system. 	
	
\begin{figure}
		\begin{center}
			\includegraphics[width=6.5in]{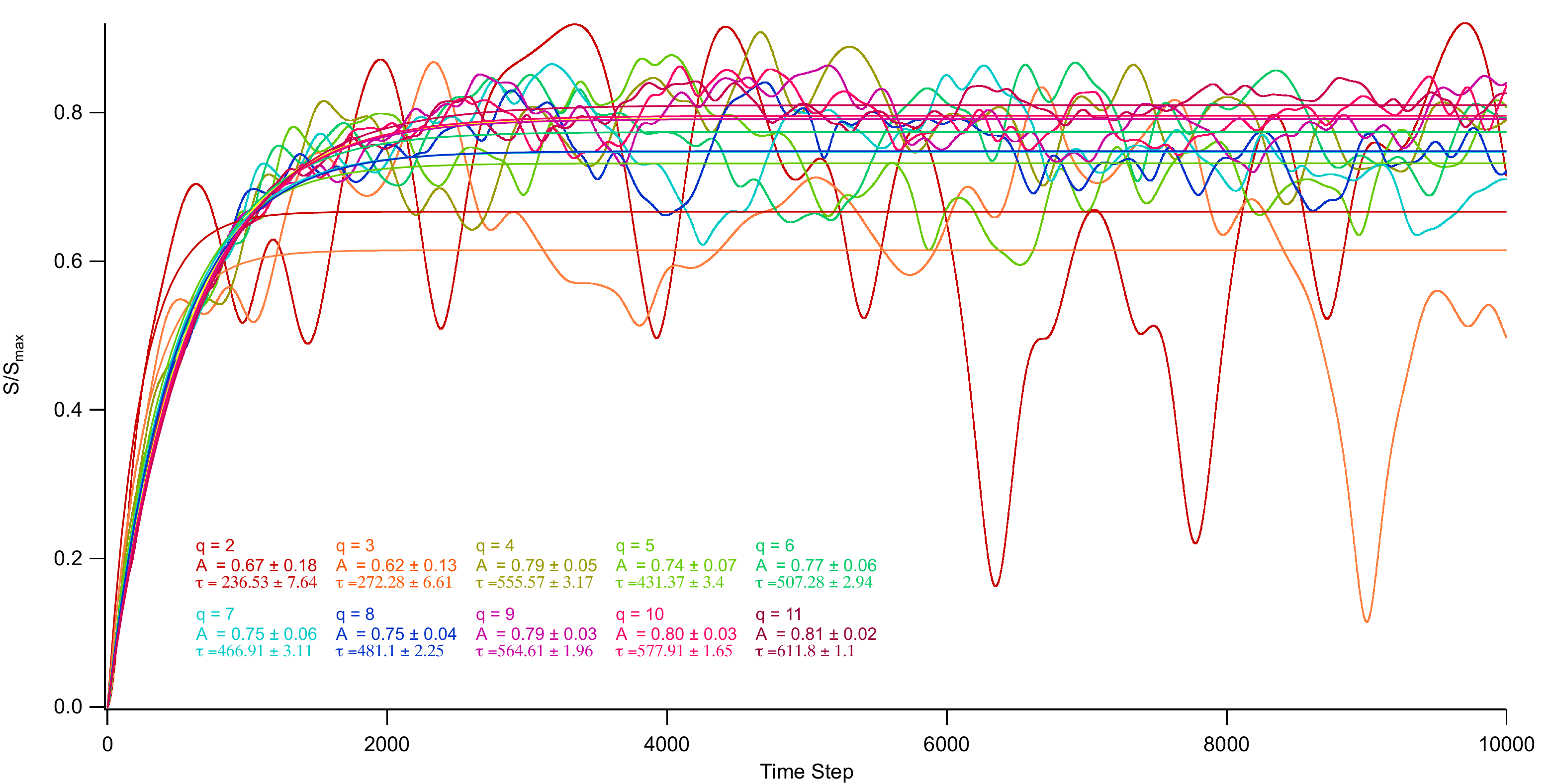} 
		\end{center}
		\caption{Von Neumann entropy fraction growth as a function of time in the Matrix theory case using a random tracing scheme. Throughout the text, the time step is set to $10^{-3} l^{(11)}_P$, and we work in Planck units where $l^{(11)}_P=1$. We generically see the entropy rising quickly -- in fact quicker for larger system sizes --- but then leveling off at an average equilibrium entropy $S_{eq}< S_{max}$. Along this equilibrium entropy plateau, the entropy fluctuates: we see larger fluctuating amplitudes for smaller system sizes as is expected from ergodic evolution. The figure also shows the results of fitting the entropy profiles with an exponential function $A (1-e^{-t/\tau})$ for different system qubits $q$: $A$ is the equilibrium entropy fraction, $A=S_{eq}/S_{max}$, and $\tau$ is the scrambling timescale in units of $t$, the simulation timestep. The resulting $A$ and $\tau$ parameters are then plotted as a function of $q$. Colors are used to tag the different $q$ cases.}\label{fig:entropygrowth}
	\end{figure}

	We run each simulation for $10^4$ time steps: we find that this time interval is long enough to generate an entropy profile that always levels off to some equilibrium value $S_{eq}$, as shown in Figure~\ref{fig:entropygrowth}. 
From each simulation, we extract two numbers: the equilibrium entropy fraction $A\equiv S_{eq}/S_{max}$, and the scrambling time $\tau$. Note that $A
\leq 1$. To determine the equilibrium entropy fraction $A$, we compute the average of $S/S_{max}$ over the range of time steps where the entropy curves fluctuates about a flat level. The standard deviation then gives us an estimate of the size of these fluctuations. To determine the scrambling time $\tau$, we employ a fitting function of the form $A (1-e^{-t/\tau})$, where $t$ is time step - with fixed $A$ computed earlier. We find that this method gives a robust estimate of scrambling time -- with associated statistical error estimation -- which qualitatively conforms well to the shape of the entropy function as can be seen from Figure~\ref{fig:entropygrowth}.
	
	We then proceed with plotting $A$ versus system qubits $q = \ln n/\ln 2$, where $n$ is the size of the system Hilbert space, $n=2^q$ -- while the value of $q$ is varied within each fixed tracing scheme. Similarly, we plot $\tau$ versus $q$\footnote{Since the proportion of system size to the whole is kept roughly constant at around $45\%$, our plots versus system qubits $q$ are qualitatively equivalent to plots versus the total number of qubits, or for that matter $N$. The horizontal axis then scales as the log of the Hilbert space size -- either the system's or the whole's.}. In all cases, our error estimation is a crude one: it mainly arises from the size of the entropy fluctuations about a plateau at late time steps. Hence, the error bars on the figures should be taken with a grain of salt. Nevertheless, we believe they do capture the relative uncertainties between the different data points on the same graph.
	  
	We next proceed with presenting all the results and conclusions from the simulations.

\subsection{Results: Matrix theory}

\begin{figure}
	\begin{center}
		\includegraphics[width=3in]{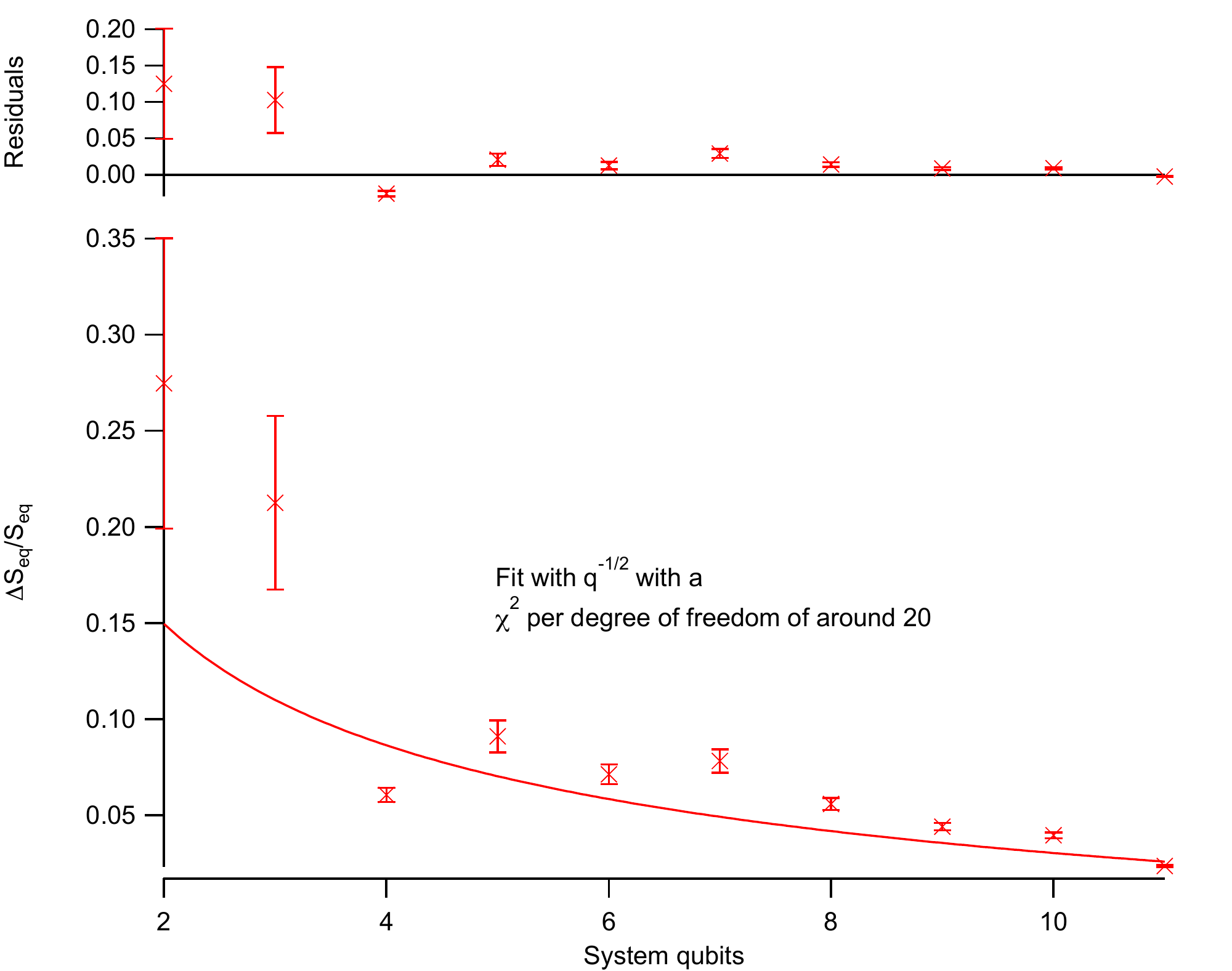} 
	\end{center}
	\caption{Entropy fluctuations and the onset of ergodicity. The case shown is that of Matrix theory with a random tracing scheme.}\label{fig:ergidicity}
\end{figure}

Figure~\ref{fig:entropygrowth} shows the evolution of the von Neumann entropy fraction as a function of simulation time step in Matrix theory employing a random tracing scheme. Similar profiles are generated for other tracing schemes. We define von Neumann entropy fraction as $A\equiv S/S_{max}$, the ratio of the entropy $S$ of the system to the maximal entropy $S_{max}=q\ln2$. The figure shows data for systems with $q=2,3,4, \ldots, 11$ qubits. These correspond to full states with respectively $5, 7, 9, \ldots, 23$ qubits in the longest chain, of which our system occupies $40\%$-$47\%$. We see from the figure that the entropy reaches an equilibrium plateau within about one thousand simulation time steps. After this, ergodicity is qualitatively obvious in the fluctuating entropy profiles, on top of which we overlay fitting curves that help us estimate the scrambling time. Fluctuations are larger for smaller systems, as expected. A closer analysis of the pattern of entropy fluctuations at equilibrium yields the expected statistical behavior for ergodic evolution: fractional fluctuations depending on the number of degrees of freedom through an inverse power law. To gauge the onset of ergodicity, Figure~\ref{fig:ergidicity} shows the fluctuations about the equilibrium entropy $S_{eq}$ for a random tracing scheme. We plot the fractional fluctuations $\Delta S_{eq}/S_{eq}$ as a function of $q$ and show a fit to $q^{-1/2}$, $q$ being the number of degrees of freedom. We see that, for $q\geq 5$, the fluctuations are roughly bounded by $10\%$. We hence consider equilibration is achieved to a desired statistical level for system qubits $q\geq 5$, or a full chain of $q\geq 11$ qubits; that is, for $N\geq 5$. For the straddled tracing scheme, similar results are achieved. However, for the contiguous tracing scheme, we find larger fluctuations -- still obeying the expected statistical pattern. For this latter case, one needs to consider a full chain with at least $q=19$ qubits to achieve ergodicity at the $10\%$ fluctuation level. This corresponds to $N\geq 10$. For all three tracing schemes, our simulations are then able to gauge equilibration.

\begin{figure}
	\begin{center}
		\includegraphics[width=6.5in]{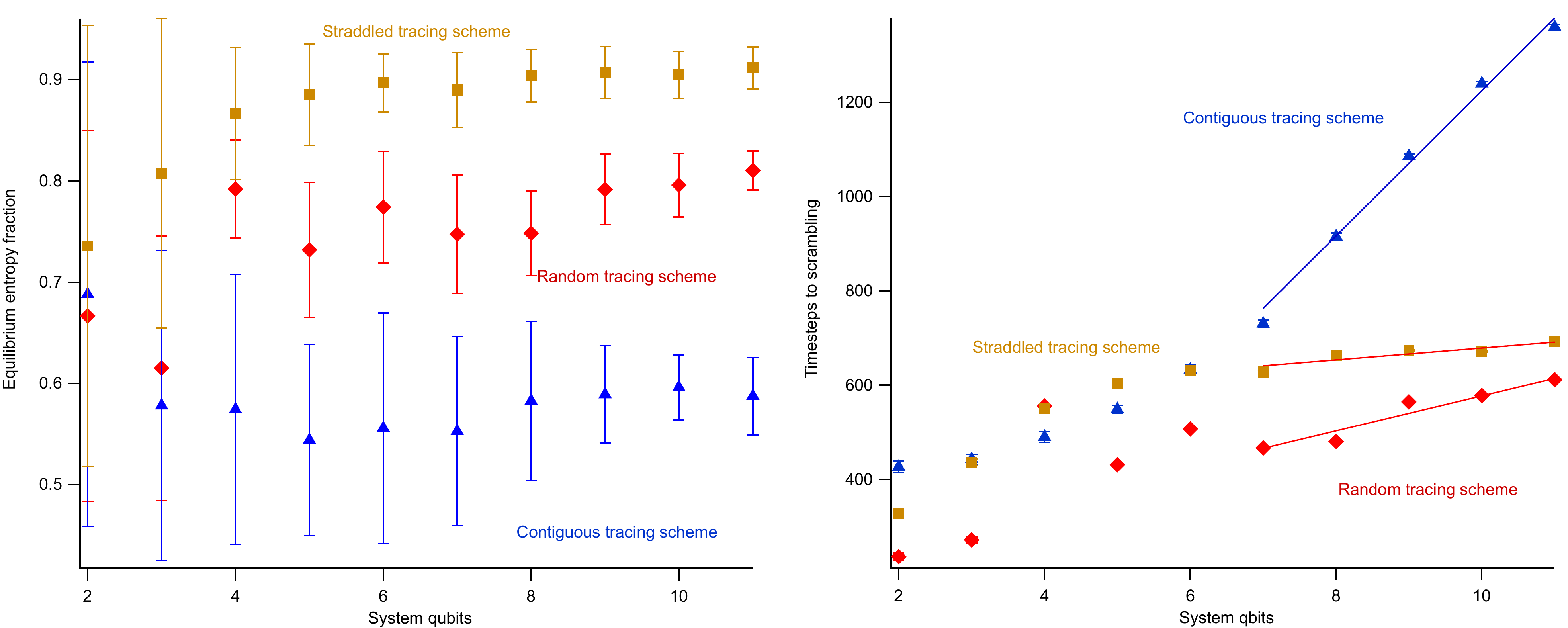} 
	\end{center}
	\caption{Equilibrium entropy fraction $A$ versus system qubits $q$ on the left and scrambling timescale $\tau$ versus $q$ on the right in Matrix theory. The horizontal axis scales as the log of the system's Hilbert space dimension $n=2^q$. We see that $\tau\sim S_{eq}$ for all three tracing schemes. The determination and estimation of the error bars are heuristic and are described in the text.}\label{fig:matrixbetaA}
\end{figure}

To fully appreciate the dynamics at work, Figure~\ref{fig:matrixbetaA} shows the results of analyzing these entropy curves, summarized in two graphs and for three different tracing schemes: a {\em straddled scheme} that maximizes the connections between the system and the whole, a {\em contiguous scheme} that minimizes the connections, and a {\em random scheme}. On the left, we show the equilibrium entropy fraction $A$ at which the system levels off, as a function of number of system qubits $q$. We notice that, as the system size grows, the equilibrium entropy fraction increases; and at around just $5$ system qubits, it reaches a $90\%$ fraction. At this point, the system is large enough to encode sufficient statistics for ergodicity: fractional entropy fluctuations beyond this point dip below $10\%$ for the straddled and random tracing schemes. For a system size with $q=5$ qubits, the whole system has $11$ qubits, and we have $U(6)$ Matrix theory. From $q=5$ to $q=11$, corresponding to Matrix theory with gauge group $U(6)$ to $U(12)$, our system is ergodic enough. We also notice that $A$ in independent $q$ beyond this point! This means that the entropy of the system scales as
\begin{equation}
	\frac{S_{eq}}{S_{max}}\sim 1 \Rightarrow S_{eq}\sim q \sim N\ .
\end{equation} 
This is the conjectured entropy scaling for a Matrix black hole~\cite{Horowitz:1997fr,Banks:1997hz,Banks:1997tn}. This issue was left unresolved in the literature since the development of the original Matrix black hole models: it was conjectured that perhaps this scaling is due to tethering D0 branes to a D2 brane membrane at the horizon. Indeed, our setup is an explicit realization of this: the spherical Matrix theory configuration we employ as background for fluctuations carries D2 brane dipole charge, and the fermionic degrees of freedom are tethered to it. And we find the conjectured and unusual $S_{eq}\sim N$ scaling. This increases our confidence that our model has the right attributes of Matrix black hole dynamics.
The second graph on the right in Figure~\ref{fig:matrixbetaA} plots scrambling time $\tau$ versus system qubits $q$. For qubits $q=5$ through $q=11$ where ergodicity is established at a $10\%$ level for the straddled and random tracing schemes, we see a clear linear dependence of $\tau$ on $q$. Remembering that the Hilbert space is of size $n=2^q$ and the equilibrium entropy we just determined scales as $S_{eq}\sim N$, the figure implies that the scrambling time scales as
\begin{equation}
	\tau\sim S_{eq}\ .
\end{equation}
To see this, we also note that a qubit flip in our system costs energy $g_s\sim R_{11}$ in tune with light-cone M theory. This means that the chemical potential for the qubit chain is of order one in Planck units, which gives a qubit chain temperature of order one as well. Hence, we have $\tau\sim S_{eq}/T\sim S_{eq}$ as noted.
Our model does not then scramble information as fast as desired in the absence of coupling to the bosonic fluctuations. This is the case even for the straddled tracing scheme which corresponds to arranging for every qubit of the system being connected to the whole -- a mechanism to generate maximal entanglement. 

\subsection{Results: BMN case}

\begin{figure}
	\begin{center}
		\includegraphics[width=6.5in]{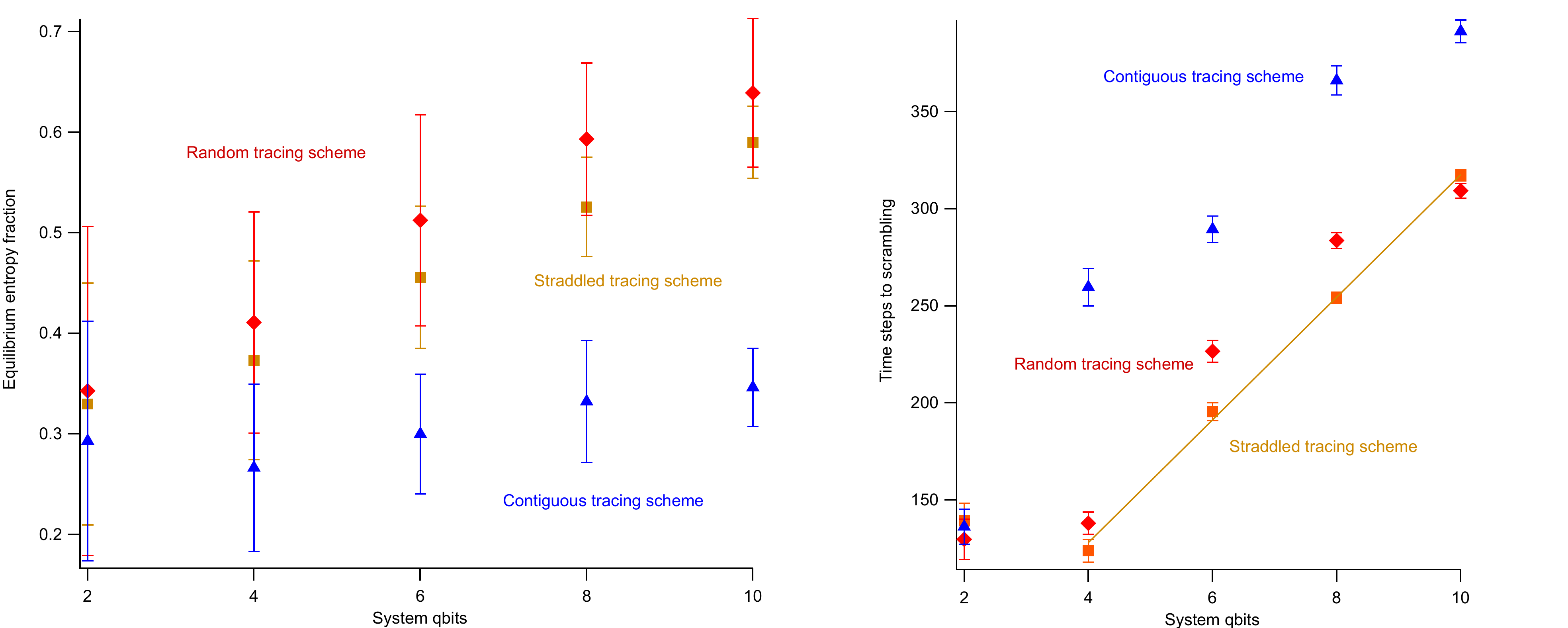} 
	\end{center}
	\caption{Equilibrium entropy fraction $A$ versus system qubits $q$ on the left and scrambling timescale $\tau$ versus $q$ on the right in the BMN model. Three tracing schemes are shown.}\label{fig:bmnA}
\end{figure}

The BMN case is a most interesting scenario because the setup is more controlled. The spherical configuration about which we perturb is a BPS state. We would then explore scrambling in a near BPS regime when $\nu=\mu/6$. However, we may expect that because of this, the evolution will be less ergodic and may require larger gauge group ranks to demonstrate the fast scrambling phenomenon if any. We find that ergodicity sets in at the $10\%$ fluctuation level for system qubits $q\geq 8$ {\em only} for the straddled tracing scheme which maximizes interactions between system and the whole. This corresponds to a full chain with $q\geq 18$ qubits, or BMN theory with $N\geq 5$.
Figure~\ref{fig:bmnA} shows the results of analyzing all three tracing schemes in the BMN model. For all three, we see that the equilibrium entropy fraction $A$ rises up with system size. This implies that the equilibrium entropy scales quadratically with $N$
\begin{equation}
	S_{eq}\sim N^2\ . 
\end{equation}
However, this conclusion is reliable only for the straddled tracing scheme, and even for that with low statistics.
The scrambling time graph seems to suggest
\begin{equation}
	\tau\sim N\sim \sqrt{S_{eq}}\ ,
\end{equation}
in particular for the straddled tracing scheme which leads to a remarkably robust linear fit in terms of $q$. We then conclude that our BMN model is not a fast scrambler when the effects of the coupling term~(\ref{eq:coupling}) are ignored. More generally, we would need to simulate the model at even higher values of $N$ to capture better statistics. However, our results are suggestive enough to provide an educated guess.

\subsection{Summary}

These results are not too surprising: the Matrix setup amounts to a one dimensional qubit chain, while the BMN case amounts to a two dimensional qubit chain. Scrambling time scaling as $S$ and $\sqrt{S}$ is very much in tune with expectations from a $d$ dimensional nearest-neighbor system, $\tau\sim S^{1/d}/T$. We conclude that the coupling to the bosonic fluctuations through~(\ref{eq:coupling}) is central to fast scrambling. The discussion in Section~\ref{sec:coupling}, analyzing the effects of such coupling, lead to very promising prospects for fast scrambling. We defer a full analysis, including an analytical treatment of the large $N$ case, to a future work~\cite{wip}.

\section{Conclusions and Outlook}\label{sec:conclusion}
\label{sub:conclusion}

In the past few years, we have come a long way from the times of the original formulation of the information paradox. In the context of string theory, the question is no more whether the evolution of probes falling into a black hole is unitary; the debate has moved instead to the details of the unitary evolution, to understanding how does the in-falling information merge with the black hole degrees of freedom. The evolution of the black hole seems to be more like the burning of a piece of paper, with no loss of information but instead the scrambling of initial data across a large Hilbert space. What may make a black hole more interesting than a piece of paper has to do with the details of the internal dynamics -- perhaps the existence of a highly efficient mechanism for scrambling information and generating entanglements at an impressive rate.

In this work, we have developed a quantum information playground in a theory of quantum gravity: the dynamics of the fermionic degrees of freedom in Matrix theory and the BMN model on a sphere. The systems were shown to consist of chains of qubits with nearest neighbor interactions in addition to a dense network of links dynamically generated by bosonic fluctuations. We demonstrated that the black hole dynamics is akin to a Brownian quantum circuit and, along with thermal back-reaction effects from couplings between the boson and fermions of the theory, the full picture looks very promising: Matrix and BMN theories appear to have the necessary ingredients to be fast scramblers. 

Can we then fully develop a complete model of a Matrix black hole -- reproducing equation of state, scrambling effect, and other special attributes of black objects? Our current setup still needs two additional ingredients.

First, we need to analytically diagonalize the Hamiltonian. We have been partially successful in doing this and we will report on the results in a separate work~\cite{wip}. This requires us to explore the scrambling dynamics in the large $N$ limit -- a desirable regime anyways. We would then write the non-unitary evolution of the open system that is the qubit chain probe using stochastic methods from quantum optics. These differential/integral equations can then be studied to identify the proper scrambling timescale in the large $N$ regime, as well as to compute thermodynamic equations of state at equilibrium. Numerical simulations which we developed in this work can then be used to test the validity of certain assumptions in such a computation.

The second ingredient has to do with the mechanism of putting the Matrix degrees of freedom in a box. In the BMN model, this is naturally provided for by the background fields and the configuration is near-BPS. In Matrix theory, we need to add by hand background curvature to confine the bosonic degrees of freedom into a spherical shape. While this is adequate to demonstrate a proof of concept fast scrambling phenomenon in the theory, it is desirable to motivate this from a more dynamical perspective. In~\cite{Sahakian:2000bg,Sahakian:2001zs}, it was suggested that such background fields may arise from back-reaction effects when $N$ is large. It was also shown in that work that, without the box, the configuration is unstable but decays to infinity parametrically slowly with larger values of $N$. Perhaps the whole story involves an accounting of the details of evaporation and incorporating the Hilbert space of the Hawking radiation in the Matrix degrees of freedom.

\begin{figure}
	\begin{center}
		\includegraphics[width=6.5in]{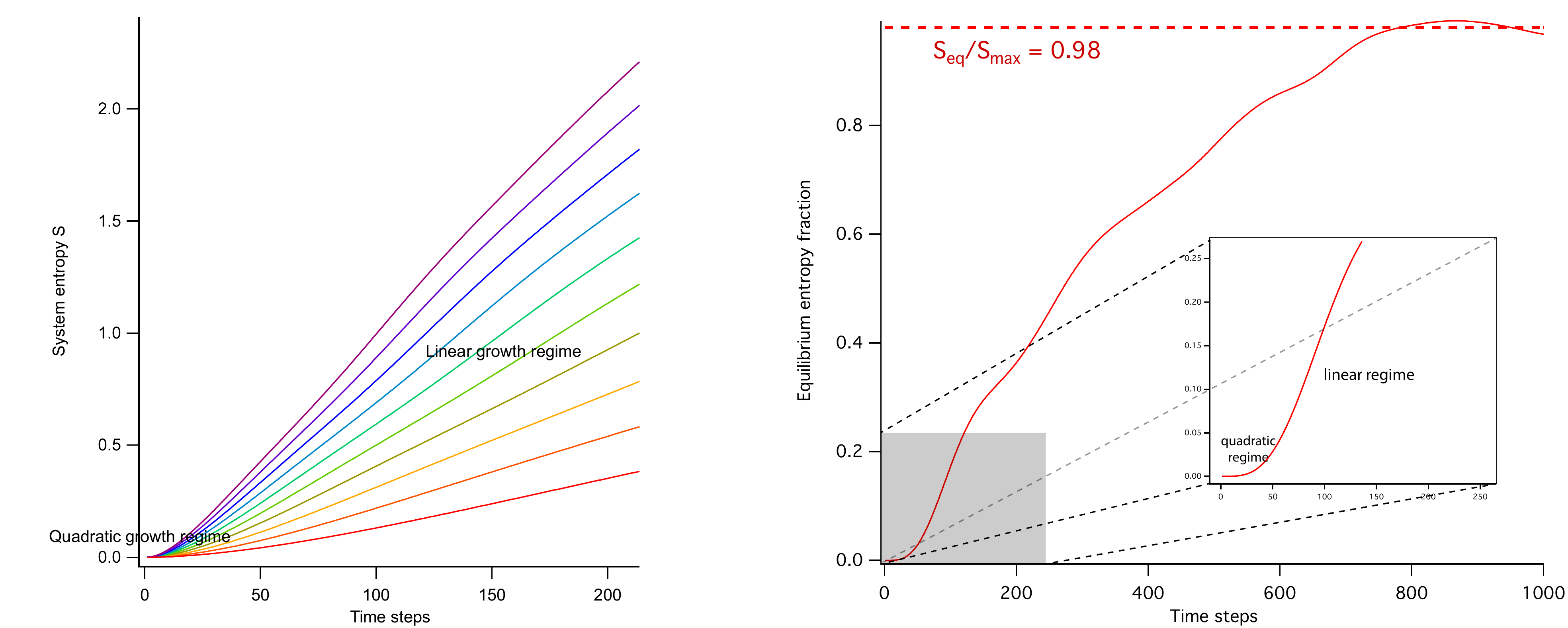} 
	\end{center}
	\caption{Early time entropy evolution in Matrix theory. On the left, we show Matrix theory with straddled tracing schemes for various qubits and in a set where the proportion of the system size to the rest is around $40$-$60$. On the right, we show the case of a system much smaller than the whole, a crude reservoir-like setup.}\label{fig:timescales}
\end{figure}

Throughout our simulations, we also noticed other timescales at work in the equilibration process, in addition to the scrambling timescale we analyzed in detail. These additional scales are most relevant at early times. Figure~\ref{fig:timescales} shows two examples. On the left, we zoom onto early entropy growth in Matrix theory simulations for various qubits while employing the straddled tracing scheme. We notice a very early growth of system entropy quadratically with time, followed by a linear growth regime. These conclusions sync well with the literature on entanglement entropy growth~\cite{Calabrese:2005in,Liu:2013iza}. To further explore this regime for setups more similar to ones investigated in the literature, we also consider a Matrix theory simulation with a system size of $3$ qubits and a whole qubit chain with $20$ qubits: this crudely approximates a system-reservoir setup, with the whole bigger than the system. Equilibration is much more robust as expected. The right graph in the figure shows the entropy growth of the system, identifying clearly an early quadratic growth regime, followed by a linear profile. Eventually, the system reaches the equilibrium entropy: the transition is smooth, as expected, since we are far from the thermodynamic limit where interesting phase transition effects may be seen.

Where does all this fit in the ongoing firewall proposal debate~\cite{Almheiri:2012rt,Almheiri:2013hfa,Harlow:2013tf,Susskind:2013tg,Maldacena:2013xja}? In a sense, our results are tangential, scrambling being relevant with or without a firewall -- the difference being the energy scale at which the scrambling will occur. The energy scale in our Matrix theory model is near the Planck scale: a qubit flip cost energy of order $g_s$, which in light-cone variables corresponds to energy of order $R_{11}$. This means that the actual energy scale of an excitation is $l_P^{(11)}$. Using equation~(\ref{eq:radius}), we are still describing a candidate large black hole of size $N\, l_P^{(11)}$ for large $N$. Put differently, the model necessarily involves a spherical D2 brane -- of Planckian substructure -- sitting at the would-be horizon, obstructing in-falling probes. Horizon dynamics in this setting is then Planckian in energy scale and favors the firewall paradigm. Given however that our model is still missing two key ingredients and hence is not a full model for a black hole, we cannot at this stage clearly and cleanly address the firewall proposal debate.

\section{Appendix}

To unravel the pattern of qubit-qubit interactions in our Matrix and BMN Hamiltonians, we choose a specific representation of the gamma matrices. Obviously the particular choice does not matter and the qubit chain structure is the same for any choice. Defining the standard Pauli matrices as
\begin{equation}
	\sigma_1 = \left(
	\begin{array}{cc}
		0	&	1	\\
		1	&	0	
	\end{array}
	\right)\ \ \ ,\ \ \ 
	\sigma_2 = \left(
	\begin{array}{cc}
		0	&	-i	\\
		i	&	0	
	\end{array}
	\right)\ \ \ ,\ \ \ 
	\sigma_3 = \left(
	\begin{array}{cc}
		1	&	0	\\
		0	&	-1	
	\end{array}
	\right)
\end{equation}
and introducing 
\begin{equation}
	\epsilon = i \sigma_2\ \ \ ,\ \ \ 1_{2\times 2} = \left(
	\begin{array}{cc}
		1	&	0	\\
		0	&	1	
	\end{array}
	\right)\ ,
\end{equation}
we set the three gamma matrices appearing in the Hamiltonians to
\begin{eqnarray}
	\gamma_1 &=& \epsilon\otimes \epsilon \otimes \epsilon\ \ \ ,\ \ \ 
	\gamma_2 = 1_{2\times 2}\otimes \sigma_1 \otimes \epsilon\ \ \ ,\ \ \ 
	\gamma_3 = 1_{2\times 2}\otimes \sigma_3 \otimes \epsilon\ .
\end{eqnarray}
A full Majorana-Weyl basis along this line can be found in~\cite{polchinski}. Our notation related to that of~\cite{polchinski} by mapping $\Gamma^7, \Gamma^8, \Gamma^9 \rightarrow \gamma_1, \gamma_2, \gamma_3$. More explicitly, this choice leads to the following $16\times 16$ matrices
\begin{equation}
	\gamma_1 = 
	\left(
	\begin{array}{cccc}
		0_{4\times 4}	&	0_{4\times 4}	&	0_{4\times 4}	&	M	\\
		0_{4\times 4}	&	0_{4\times 4}	&	-M	&	0_{4\times 4}	\\
		0_{4\times 4}	&	-M	&	0_{4\times 4}	&	0_{4\times 4}	\\
		M	&	0_{4\times 4}	&	0_{4\times 4}	&	0_{4\times 4}	
	\end{array}
	\right)\ \ \ ,\ \ \ 
	\gamma_2 = 
	\left(
	\begin{array}{cccc}
		0_{4\times 4}	&	0_{4\times 4}	&	1_{4\times 4}	&	0_{4\times 4}	\\
		0_{4\times 4}	&	0_{4\times 4}	&	0_{4\times 4}	&	1_{4\times 4}	\\
		1_{4\times 4}	&	0_{4\times 4}	&	0_{4\times 4}	&	0_{4\times 4}	\\
		0_{4\times 4}	&	1_{4\times 4}	&	0_{4\times 4}	&	0_{4\times 4}	
	\end{array}
	\right) 
\end{equation}
\begin{equation}
	\gamma_3 = 
	\left(
	\begin{array}{cccc}
		-1_{4\times 4}	&	0_{4\times 4}	&	0_{4\times 4}	&	0_{4\times 4}	\\
		0_{4\times 4}	&	-1_{4\times 4}	&	0_{4\times 4}	&	0_{4\times 4}	\\
		0_{4\times 4}	&	0_{4\times 4}	&	1_{4\times 4}	&	0_{4\times 4}	\\
		0_{4\times 4}	&	0_{4\times 4}	&	0_{4\times 4}	&	1_{4\times 4}	
	\end{array}
	\right) 
\end{equation}
where we define
\begin{equation}
	M \equiv 
	\left(
	\begin{array}{cccc}
		-1	&	0	&	0	&	0	\\
		0	&	+1	&	0	&	0	\\
		0	&	0	&	-1	&	0	\\
		0	&	0	&	0	&	+1
	\end{array}
	\right)\ .
\end{equation}

\section{Acknowledgments}

This work was supported by NSF grant number PHY-0968726, and a gift of two Tesla GPU cards from the nVidia Corporation.


\providecommand{\href}[2]{#2}\begingroup\raggedright\endgroup

\end{document}